\documentclass[draftcls,onecolumn,11pt]{IEEEtran}

\usepackage{amssymb,latexsym,color,fancybox,graphicx,mathrsfs,url,coolstr,bm,amsthm}
\usepackage[normalem]{ulem}
\usepackage{multimedia}
\usepackage{multirow}
\usepackage{rotating}
\usepackage{epstopdf}
\usepackage{geometry}
\usepackage{cite}
\graphicspath{{./Figures/}}

\renewcommand{\[}{\begin{equation}}
\renewcommand{\]}[1]{\label{eq:#1}\end{equation}}

\newcommand{\bfig}{\begin{figure}[p]}
\newcommand{\efig}[1]{\label{fig:#1}\end{figure}}

\newcommand{\btab}{\begin{table}[p]}
\newcommand{\etab}[1]{\label{tab:#1}\end{table}}

\definecolor{deep-blue}{rgb}{0.5,0.53,0.75}
\definecolor{deep-red}{rgb}{0.7,0.3,0.3}
\definecolor{deeper-blue}{rgb}{0.25,0.28,0.5}

 \newtheorem{theorem}{Theorem}

 \newtheorem{lemma}{Lemma}
 \newtheorem{corollary}{Corollary}

\renewcommand{\P}[1]{p\left(#1\right)}

\newcommand{\sign}{\mathop{\rm sign}}
\renewcommand{\c}{{\mathbf{c}}}

\newcommand{\etal}{\emph{et\,al. }}

\newcommand{\ttheta}{{\bm\theta}}

\renewcommand{\u}{\mathbf u}

\renewcommand{\a}{{\mathbf{a}}}
\newcommand{\x}{{\mathbf{x}}}
\newcommand{\hx}{{\hat{\mathbf{x}}}}
\newcommand{\y}{{\mathbf{y}}}

\newcommand{\f}{{\mathbf{f}}}
\newcommand{\p}{{\mathbf{p}}}

\renewcommand{\u}{{\mathbf{u}}}
\newcommand{\E}[1]{{\mathscr E}\left\{#1\right\}}

\newcommand{\Dop}[1]{{\mathscr D}\setbox0=\hbox{#1}\ifdim\wd0=0ex\else\left\{#1\right\}\fi}
\newcommand{\Rop}[1]{{\mathscr R}\setbox0=\hbox{#1}\ifdim\wd0=0ex\else\left\{#1\right\}\fi}
\newcommand{\D}{{\mathbf{D}}}

\newcommand{\R}{{\mathbf{R}}}
\newcommand{\I}{{\mathbf{Id}}}

\newcommand{\M}{{\mathbf{M}}}

\newcommand{\w}{{\mathbf{w}}}
\newcommand{\1}{{\mathbf{1}}}
\newcommand{\4}{{\mathbf{4}}}

\newcommand{\tD}{{\overline{\mathbf{D}}}}
\newcommand{\ttw}{{\overline{\mathbf{w}}}}
\newcommand{\tw}{{\overline{w}}}

\newcommand{\transp}{^{\textsc{\tiny T}}}

\newcommand{\var}[1]{{\mathop{\rm var}\left\{#1\right\}}}

\newcommand{\dB}{\,\mbox{dB}}

\newcommand{\ds}{\displaystyle}

\newcommand{\bs}{\bm}
\newcommand{\m}{{\mathbf{m}}}
\newcommand{\s}{{\mathbf{s}}}

\newcommand{\re}[1]{{\mathop{\Re}\left\{#1\right\}}}
\newcommand{\im}[1]{{\mathop{\Im}\left\{#1\right\}}}

\newcommand{\oomega}{\bs\omega}
\newcommand{\mmu}{\bs\mu}
\newcommand{\vvarsigma}{\bs\varsigma}

\newcommand{\MSE}{\mbox{MSE}}

\newcommand{\SNR}{\mbox{SNR}}
\newcommand{\PSNR}{\mbox{PSNR}}
\newcommand{\SSIM}{\mbox{SSIM}}
\newcommand{\CIPSNR}{\mbox{CIPSNR}}

\newcommand{\filtre}[2]{\begin{picture}(45,10)
\put(0,5){\line(1,0){10}}
\put(10,0){\framebox(35,10){$#1$}}
\put(45,5){\line(1,0){10}}
\put(55,5){\makebox(0,0)[l]{\ $#2$}}
\end{picture}}
\newcommand{\filtreA}[3]
{\begin{picture}(55,10)
	\put(0,5){\line(1,0){7}}
	\put(7,0){\framebox(21,10){$#1$}}
	\put(28,5){\line(1,0){7}}
	\put(40,5){\circle{10}}
	\put(40,5){\makebox(0,0){$\downarrow\!#2$}}
	\put(45,5){\line(1,0){10}}
	\put(55,6){\makebox(0,0)[l]{$#3$}}
\end{picture}}
\newcommand{\filtreS}[3]
{\begin{picture}(55,10)
	\put(0,5){\line(1,0){10}}
	\put(15,5){\circle{10}}
	\put(15,5){\makebox(0,0){$\uparrow\!#2$}}
	\put(20,5){\line(1,0){7}}
	\put(27,0){\framebox(21,10){$#1$}}
	\put(48,5){\line(1,0){7}}
	\put(55,6){\makebox(0,0)[l]{$#3$}}
\end{picture}}

\begin{document}
\title{A CURE for Noisy Magnetic Resonance Images: Chi-Square Unbiased Risk Estimation}
\author{Florian Luisier\thanks{Florian Luisier and Patrick J. Wolfe are with the Statistics and Information Sciences Laboratory, Harvard University, Cambridge, MA 02138, USA (email: fluisier@seas.harvard.edu, wolfe@stat.harvard.edu)}, Thierry Blu\thanks{Thierry Blu is with the Department of Electronic Engineering, The Chinese University of Hong Kong, Shatin, N.T., Hong Kong (e-mail: thierry.blu@m4x.org).}, and Patrick J. Wolfe
\thanks{This work was supported by the Swiss National Science Foundation Fellowship BELP2-133245, the US National Science Foundation Grant~DMS-0652743, and the General Research Fund CUHK410209 from the Hong Kong Research Grant Council.}\vspace{-2\baselineskip}}

\markboth{Submitted Manuscript}{Chi-Square Unbiased Risk Estimate for Magnitude MR Image Denoising}
\maketitle


\IEEEpeerreviewmaketitle

\begin{abstract}
In this article we derive an unbiased expression for the expected mean-squared error associated with continuously differentiable estimators of the noncentrality parameter of a chi-square random variable.  We then consider the task of denoising squared-magnitude magnetic resonance image data, which are well modeled as independent noncentral chi-square random variables on two degrees of freedom.  We consider two broad classes of linearly parameterized shrinkage estimators that can be optimized using our risk estimate, one in the general context of undecimated filterbank transforms, and another in the specific case of the unnormalized Haar wavelet transform.  The resultant algorithms are computationally tractable and improve upon state-of-the-art methods for both simulated and actual magnetic resonance image data.
\end{abstract}

\begin{center}
	\bfseries EDICS: TEC-RST (primary);  COI-MRI, SMR-SMD, TEC-MRS (secondary)
\end{center}

\baselineskip=23pt

\section{Introduction}
\label{sec:Intro}
Magnetic resonance (MR) imaging is a fundamental \emph{in vivo} medical imaging technique that provides high-contrast images of soft tissue without the use of ionization radiation.  The signal-to-noise ratio (SNR) of an acquired MR image is determined by numerous physical and structural factors, such as static field strength, resolution, receiver bandwidth, and the number of signal averages collected at each encoding step~\cite{Wright1997,Epstein2008}. Current developments in MR imaging are focused primarily on lowering its inherently high scanning time, increasing the spatiotemporal resolution of the images themselves, and reducing the cost of the overall system. However, the pursuit of any of these objectives has a negative impact on the SNR of the acquired image. A post-acquisition denoising step is therefore essential to clinician visualization and meaningful computer-aided diagnoses~\cite{Wright1997}.

In MR image acquisition, the data consist of discrete Fourier samples, usually referred to as $k$-space samples.  These data are corrupted by random noise, due primarily to thermal fluctuations generated by a patient's body in the imager's receiver coils~\cite{Wright1997,Epstein2008}. This random degradation is well modeled by additive white Gaussian noise (AWGN) that independently corrupts the real and imaginary parts of the complex-valued $k$-space samples~\cite{Henkelman1985}. Assuming a Cartesian sampling pattern, the output image is straightforwardly obtained by computing the inverse discrete Fourier transform of the $k$-space samples; the resolution of the reconstructed image is then determined by the maximum $k$-space sampling frequency. Note that some recent works (see~\cite{Lustig2008} and references therein) aim to accelerate MR image acquisition time by undersampling the $k$-space. Non-Cartesian (e.g., spiral or random) sampling trajectories, as well as nonlinear (usually sparsity-driven) reconstruction schemes are then considered. This axis of research is, however, outside the scope of the present work.

Following application of an inverse discrete Fourier transform, the resulting image may be considered as a complex-valued signal corrupted by independent and identically distributed samples of complex AWGN. In magnitude MR imaging, the image phase is disregarded and only the magnitude is considered for visualization and further analysis. Although the samples of the magnitude image remain statistically independent, they are no longer Gaussian, but rather are Rician distributed~\cite{Gudbjartsson1995}. Contrary to the Gaussian case, this form of ``noise'' is \emph{signal-dependent} in that both the mean and the variance of the magnitude samples depend on the underlying noise-free magnitude image. Consequently, generic denoising algorithms designed for AWGN reduction usually do not give satisfying results on Rician image data.

Denoising of magnitude MR images has thus gained much attention over the past several years. Two main strategies can be distinguished as follows. In the first, the Rician data are treated directly, often in the image domain. In the second, the denoising is applied to the squared magnitude MR image, which follows a (scaled) noncentral chi-square distribution on two degrees of freedom, whose noncentrality parameter is proportional to the underlying noise-free squared magnitude.  An appealing aspect of this strategy is that it renders the bias due to the noise constant rather than signal dependent, thus enabling standard approaches and transform-domain shrinkage estimators to be applied.

Following the first strategy previously mentioned, several Rician-based maximum likelihood estimators~\cite{Sijbers1998b,Sijbers2004,He2009} have been proposed. In~\cite{He2009}, this estimation is performed non-locally among pixels having a similar neighborhood. A Bayesian maximum a posteriori estimator has been devised in~\cite{Awate2007}, with the prior modeled in a nonparametric Markov random field framework. Very recently, Foi proposed a Rician-based variance stabilizing transform which makes the use of standard AWGN denoisers more effective~\cite{Foi2011}. Following the second strategy, a linear minimum mean-squared error filter applied to the squared magnitude image has been derived in~\cite{Fernandez2008a,Fernandez2008b}.  In addition to these statistical model-based approaches, much work has also been devoted to the adaptation and enhancement of the nonlocal means filter originally developed by Buades \etal for AWGN reduction~\cite{Buades2005}. The core of this relatively simple, yet effective, denoising approach consists of a weighted averaging of similarly close (in spatial and photometric distance) pixels. Some of these adaptations operate directly on the Rician magnitude image data~\cite{Coupe2008b,Gal2010}, while others are applied to the squared magnitude image~\cite{Manjon2008,Liu2010,Thaipanich2010}.

In addition to these image-domain approaches, several magnitude MR image denoising algorithms have also been developed for the wavelet domain.  The sparsifying and decorrelating effects of wavelets and other related transforms typically result in the concentration of relevant image features into a few significant wavelet coefficients. Simple thresholding rules based on coefficient magnitude then provide an effective means of reducing the noise level while preserving sharp edges in the image.  In the earliest uses of the wavelet transform for MR image denoising~\cite{Weaver1991,Xu1994}, the Rician distribution of the data was not explicitly taken into account.  Nowak subsequently proposed wavelet coefficient thresholding based on the observation that the empirical wavelet coefficients of the squared-magnitude data are unbiased estimators of the coefficients of the underlying squared-magnitude image, and that the residual scaling coefficients exhibit a signal-independent bias that is easily removable~\cite{Nowak1999}.  While the pointwise coefficient thresholding proposed in~\cite{Nowak1999} is most natural in the context of an orthogonal discrete wavelet transform, Pi\v{z}urica \etal subsequently developed a Bayesian wavelet thresholding algorithm applied in an undecimated wavelet representation~\cite{Pizurica2003}.  Wavelet-based denoising algorithms that require the entirety of the \emph{complex} MR image data~\cite{Wood1999,Zaroubi2000,Bao2003} are typically applied separately to the real and imaginary components of the image.

In this article, we develop a general result for chi-square unbiased risk estimation, which we then apply to the task of denoising squared-magnitude MR images.  We provide two instances of effective transform-domain algorithms, each based on the concept of \emph{linear expansion of thresholds} (LET) introduced in~\cite{Luisier2007a,Blu2007}. The first class of proposed algorithms consists of a pointwise continuously differentiable thresholding applied to the coefficients of an undecimated filterbank transform. The second class takes advantage of the conservation of the chi-square statistics across the lowpass channel of the unnormalized Haar wavelet transform. Owing to the remarkable property of this orthogonal transform, it is possible to derive independent risk estimates in each wavelet subband, allowing for a very fast denoising procedure. These estimates are then used to optimize the parameters of subband-dependent joint inter-/intra-scale LET.

The article is organized as follows.  We first derive in Section~\ref{sec:CURE} an unbiased expression of the risk associated with estimators of the noncentrality parameter of a chi-square random variable having arbitrary (known) degrees of freedom.  Then, in Section~\ref{sec:CUREuwt}, we apply this result to optimize pointwise estimators for undecimated filterbank transform coefficients, using linear expansion of thresholds.  In Section~\ref{sec:CUREhaar}, we give an expression for chi-square unbiased risk estimation directly in the unnormalized Haar wavelet domain, and propose a more sophisticated joint inter-/intra-scale LET.  We conclude in Section~\ref{sec:Experiments} with denoising experiments conducted on simulated and actual magnitude MR images, and evaluations of our methods relative to the current state of the art.  Note that a subset of this work (mainly part of Section~\ref{sec:CUREuwt}) has been accepted for presentation at the 2011 IEEE International Conference on Image Processing~\cite{Luisier2011b}.

\section{A Chi-Square Unbiased Risk Estimate (CURE)}
\label{sec:CURE}
Assume the observation of a vector  $\y\in\mathbb{R}^N_{+}$ of $N$ independent samples $y_n$, each randomly drawn from a noncentral chi-square distribution with (unknown) noncentrality parameter $x_n \geq 0$ and (known) common degrees of freedom $K > 0$.  We use the vector notation $\y\sim\chi^2_K(\x)$, recalling that (for integer $K$) the joint distribution $\P{\y|\x}$ can be seen as resulting from the addition of $K$ independent vectors on $\mathbb{R}^N_+$ whose coordinates are the \emph{squares} of non-centered Gaussian random variables of unit variance. This observation model is statistically characterized by the data likelihood
\begin{equation}
\label{eq:NC2}
\P{\y|\x}=\prod^N_{n=1}\P{y_n|x_n} =\prod^N_{n=1}\frac{1}{2}\mathrm{e}^{-\frac{x_n+y_n}{2}}\left(\frac{y_n}{x_n}\right)^{\frac{K-2}{4}}I_{\frac{K}{2}-1}(\sqrt{x_n y_n}) \text{,}
\end{equation}
where $\ds I_{\alpha}(u)=\sum_{k\in\mathbb{N}}\frac{1}{k!\Gamma(k+\alpha+1)}\left(\frac{u}{2}\right)^{2k+\alpha}$ is the $\alpha$-order modified Bessel function of the first kind.

The chi-square distribution of~\eqref{eq:NC2} is most easily understood through its characteristic function, or Fourier transform:
\begin{equation}
\label{eq:FourierNC2}
\hat p(\oomega|\x)=\prod_{n=1}^N\frac{\displaystyle\exp\Bigl(-\frac{j\omega_nx_n}{1+2j\omega_n}\Bigr)}{\displaystyle(1+2j\omega_n)^{K/2}}
\text{.}
\end{equation}
For instance, by equating first and second order Taylor developments of this Fourier transform with the corresponding moments, it is straightforward to show that
\begin{equation}
\label{eq:momentsNC2}
\begin{array}{l}
  \ds \hspace*{1.25em}\E{\,\y\,}=\x+K\cdot\1,  \\
  \ds\makebox[0em][r]{ and \ }\E{\|\y\|^2}=\|\x\|^2+2(K+2)\1\transp\x+NK(K+2) \text{,}\\
\end{array}
\end{equation}
where $\E{\cdot}$ is the expectation operator.

When designing an estimator $\hx=\f(\y)$ of $\x$, a natural criterion is the value of its associated risk or expected mean-squared error (MSE), defined here by
\begin{eqnarray}
\label{eq:MSE}
\E{\MSE}
=
\E{\frac{1}{N}\Vert\f(\y)-\x\Vert^2}
=
\frac{1}{N}\sum^N_{n=1}\left(\E{f_n(\y)^2}-2\E{x_nf_n(\y)}+x^2_n\right) \text{.}
\end{eqnarray}
The expectation in~\eqref{eq:MSE} is with respect to the data $\y$; the vector $\x$ of noncentrality parameters may either be considered as deterministic and unknown, or as random and independent of $\y$.

In practice, for any given realization of the data $\y$, the true MSE cannot be computed because $\x$ is unknown. Yet, paralleling the general case of distributions in the exponential family~\cite{Brown1986}, it is possible to establish a lemma that allows us to circumvent this issue, and \emph{estimate} the risk without knowledge of $\x$.  The main technical requirement is that each component $f_n(\y)$ of the estimator $\f:\mathbb{R}^N\rightarrow\mathbb{R}^N$ be continuously differentiable with respect to $y_n$, and so we introduce the following notation:
$\partial\f(\y) = \left[\frac{\partial}{\partial y_n}f_n(\y)\right]_{1\leq n \leq N}$, $\partial^2\f(\y) = \left[\frac{\partial^2}{\partial y^2_n}f_n(\y)\right]_{1\leq n \leq N}$.

\begin{lemma}
\label{lem:CURE}
Assume that the estimator $\f(\y)=[f_n(\y)]_{1\leq n \leq N}$ is such that each $f_n(\y)$ is continuously differentiable with respect to $y_n$, with weakly differentiable partial derivatives $\partial f_n(\y) / \partial y_n$ that do not increase ``too quickly'' for large values of $\y$; i.e., there exists a constant $s<1/2$ such that for every $1\leq n \leq N$,
$
\lim_{y_n\to+\infty}\mathrm{e}^{-sy_n}|\frac{\partial}{\partial y_n}f_n(\y)|=0
\text{.}
$
Then
\begin{equation}
\label{eq:CURElem}
\E{\x\transp\f(\y)} =\E{(\y-K\cdot\1)\transp\f(\y)}-4\E{\left(\y-\frac{K}{2}\cdot\1\right)\transp\partial\f(\y)}+4\,\E{\y\transp\partial^2\f(\y)}
\text{.}
\end{equation}
\end{lemma}
\begin{IEEEproof}
We first evaluate the scalar expectation $\E{x_nf_n(\y)}$ appearing in~\eqref{eq:MSE}, before summing up the contributions over $n$ to get the expectation of the inner product $\x\transp\f(\y)$.

Let us consider the characteristic function given by~\eqref{eq:FourierNC2}. By differentiating the logarithm of this Fourier transform w.r.t. the variable $\omega_n$, we find that
$$
\frac{1}{\hat p(\oomega|\x)}\frac{\partial \hat p(\oomega|\x)}{\partial\omega_n}=-\frac{jK}{1+2j\omega_n}-\frac{jx_n}{(1+2j\omega_n)^2} \text{.}
$$
After a rearrangement of the different terms involved, we get
$$
x_n\hat p(\oomega|\x)=j(1+2j\omega_n)^2\frac{\partial \hat p(\oomega|\x)}{\partial\omega_n}-K(1+2j\omega_n)\hat p(\oomega|\x) \text{.}
$$
Using $(1+2j\omega_n)^2=1+4j\omega_n+4(j\omega_n)^2$ and recalling that multiplication by $j\omega_n$ corresponds to differentiating w.r.t. $y_n$, while differentiation w.r.t. $\omega_n$ is equivalent to a multiplication with $-jy_n$, we can deduce that the probability density $p(\y|\x)$ satisfies (in the sense of distributions) the following linear differential equation
$$
x_np(\y|\x)=\Bigl(1+4\frac{\partial}{\partial y_n}+4\frac{\partial^2}{\partial y_n^2}\Bigr)\bigl\{y_np(\y|\x)\bigr\}
-K\Bigl(1+2\frac{\partial}{\partial y_n}\Bigr)\bigl\{p(\y|\x)\bigr\}.
$$

The expectation $\E{x_nf_n(\y)}$ is simply the Euclidean inner product (with respect to Lebesgue measure) between $f_n(\y)$ and $x_np(\y|\x)$.  Continuous differentiability of $\f(\y)$ coupled with weak differentiability of $\partial\f(\y)$ implies that we may twice apply integration by parts to obtain
$$
\begin{array}{l}
    \ds \E{x_nf_n(\y)} =\langle f_n(\y),x_np(\y|\x)\rangle \\
    \ds\phantom{\E{x_nf_n(\y)}}   = \Bigl\langle f_n(\y),\Bigl(1+4\frac{\partial}{\partial y_n}+4\frac{\partial^2}{\partial y_n^2}\Bigr)\bigl\{y_np(\y|\x)\bigr\} \Bigr\rangle
    -K \Bigl\langle f_n(\y),\Bigl(1+2\frac{\partial}{\partial y_n}\Bigr)\bigl\{p(\y|\x)\bigr\} \Bigr\rangle\\
    \ds\phantom{\E{x_nf_n(\y)}}   = \Bigl\langle y_n\Bigl(1-4\frac{\partial}{\partial y_n}+4\frac{\partial^2}{\partial y_n^2}\Bigr)\bigl\{f_n(\y)\bigr\},p(\y|\x) \Bigr\rangle
-K \Bigl\langle \Bigl(1-2\frac{\partial}{\partial y_n}\Bigr)\bigl\{f_n(\y)\bigr\},p(\y|\x) \Bigr\rangle\\
    \ds\phantom{\E{x_nf_n(\y)}}   = \E{ (y_n-K)f_n(\y)-4\left(y_n-\frac{K}{2}\right)\frac{\partial f_n(\y)}{\partial y_n}+4y_n\frac{\partial^2f_n(\y)}{\partial y_n^2}}
    \end{array}
$$
plus additional integrated terms that do not depend on $x_n$, which vanish because of our (conservative) assumption on how $f_n(\y)$ increases when $y_n\to+\infty$.  (Asymptotic expansions of Bessel functions can be used to show that whenever $s<1/2$, we have that $\mathrm{e}^{sy_n}\P{\y|\x}\to0$ and $\mathrm{e}^{sy_n}\partial\P{\y|\x}/\partial y_n\to0$.)  Finally, summing up over the index $n$ yields~\eqref{eq:CURElem}.
\end{IEEEproof}

We can then deduce a theorem that provides an unbiased estimate of the expected MSE given in~\eqref{eq:MSE}. We term this random variable CURE, an acronym for Chi-square Unbiased Risk Estimate.
\begin{theorem}
\label{th:CURE}
Let $\y\sim\chi^2_K(\x)$ and assume that $\f(\y)$ satisfies the regularity conditions of Lemma~\ref{lem:CURE}. Then, the following random variable:
\begin{eqnarray}
\label{eq:CURE}
\emph{CURE} =
\frac{1}{N}\left(\Vert \f(\y)\!-\!(\y\!-\!K\!\cdot\!\1)\Vert^2\!-\!\4\transp\left(\y\!-\!\frac{K}{2}\!\cdot\!\1\right)\right)+
\frac{8}{N}\left(\left(\y\!-\!\frac{K}{2}\!\cdot\!\1\right)\transp\partial\f(\y)\!-\!\y\transp\partial^2\f(\y)\right)
\text{,}
\end{eqnarray}
is an unbiased estimate of the risk; i.e., $\E{\emph{CURE}}=\E{\emph{MSE}}$.
\end{theorem}
\begin{IEEEproof}
As with other unbiased risk estimates, we express the MSE as a sum of three terms:
$$
\|\f(\y)-\x\|^2=\underbrace{\|\f(\y)\|^2}_{{\rm term}\ 1}-\underbrace{2\x\transp\f(\y)}_{{\rm term}\ 2}+\underbrace{\|\x\|^2}_{{\rm term}\ 3} \text{,}
$$
which we replace by a statistical equivalent \emph{that does not depend on $\x$ anymore}.

Term~$1$ needs no change, term~$2$ can be replaced according to~\eqref{eq:CURElem} from Lemma~\ref{lem:CURE}, and term~3 can be reformulated using the noncentral chi-square moments of~\eqref{eq:momentsNC2} to yield
$$
\|\x\|^2=\E{\|\y\|^2}-2(K+2)\1\transp\E{\y}+NK(K+2).
$$
Putting everything together then leads directly to~\eqref{eq:CURE}, and thus the theorem is proved.
\end{IEEEproof}

In Sections~\ref{sec:CUREuwt} and~\ref{sec:CUREhaar} we propose specific examples of CURE-optimized transform-domain processing for estimating the unknown noncentrality parameter vector $\x$ from data $\y$.

\section{CURE-Optimized Denoising in Undecimated Filterbank Transforms}
\label{sec:CUREuwt}
In this section, we focus on processing within a $(J+1)$-band undecimated filterbank transform, as depicted in Fig.~\ref{fig:UFT}. This broad class of redundant representations notably includes the undecimated wavelet transform (UWT) and overlapping block discrete cosine transform (BDCT).
\thinlines
\setlength{\unitlength}{0.6mm}
\begin{figure}[h!]
\centering
\begin{picture}(155,45)
\put(0,31){\makebox(0,0)[r]{$\y$}}
\put(0,31){\line(1,0){10}}
\put(10,21){\line(0,1){30}}
\put(10,5){\line(0,1){5}}
\put(10,10){\makebox(0,16){\vdots}}
\put(10,0){\filtre{\tilde G_{J}(z^{-1})}{\w^{J}}}
\put(10,-3){$\underbrace{\phantom{\hspace{33mm}}}_{}$}
\put(10,-15){$\D=[\D\transp_0\:\:\D\transp_1\ldots\D\transp_J]\transp$}
\put(70,10){\makebox(0,16){\vdots}}
\put(10,26){\filtre{\tilde G_1(z^{-1})}{\w^1}}
\put(10,46){\filtre{\tilde G_0(z^{-1})}{\w^0}}
\put(77,0){\filtre{G_{J}(z)}{}}
\put(77,-3){$\ds\underbrace{\phantom{\hspace{33mm}}}_{}$}
\put(77,-15){$\R=[\R_0\:\:\R_1\ldots\R_J]$}
\put(77,26){\filtre{G_1(z)}{}}
\put(77,46){\filtre{G_0(z)}{}}
\put(142,31){\makebox(0,0)[l]{$\,\y$}}
\put(132,31){\makebox(0,0){$\bigoplus$}}
\put(132,31){\line(1,0){10}}
\put(132,21){\line(0,1){30}}
\put(132,5){\line(0,1){5}}
\put(132,10){\makebox(0,16){\vdots}}
\end{picture}
\vspace{1.5em}
\caption{Undecimated $(J+1)$-band analysis/synthesis filterbank.}
\label{fig:UFT}
\vspace{-2em}
\end{figure}
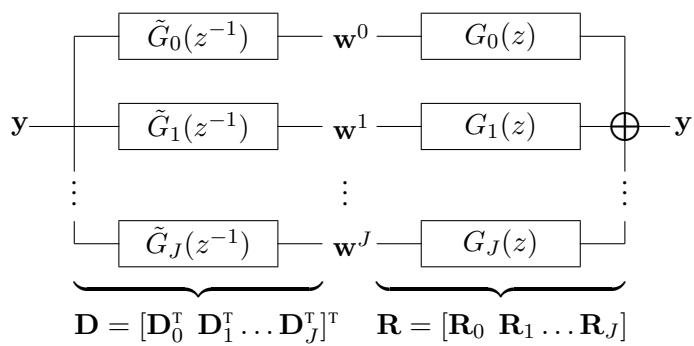

\subsection{Image-domain CURE for transform-domain processing}
\label{ssec:CUREuwt}
Nonlinear processing via undecimated filterbank transforms is a general denoising strategy long proven to be effective for reducing various types of noise degradations \cite{Coifman1995,Nason1995,Gyaourova2002,Starck2007,Blu2007,Luisier2011a}. It essentially boils down to performing a linear (and possibly redundant) analysis transformation of the data, which provides empirical coefficients that are then thresholded (possibly using a multivariate nonlinear function), the result of which is finally passed to a linear synthesis transformation. When treating signal-dependent noise, the entire denoising procedure is conveniently  expressed in the generic form $\f(\y)=\R\ttheta(\D\y,\tD\y)$~\cite{Luisier2011a}, where
\begin{itemize}
\item The circulant matrices $\D=[\D\transp_0\:\:\D\transp_1\ldots\D\transp_J]\transp$ and $\R=[\R_0\:\:\R_1\ldots\R_J]$ implement an arbitrary pair of analysis/synthesis undecimated filterbanks (Fig.~\ref{fig:UFT}) such that the perfect reconstruction condition $\R\D=\I$ is satisfied. Each component of the $N\times N$ circulant submatrix $\D_j=[d^j_{k,l}]_{1\leq k,l\leq N}$ and $\R_j=[r^j_{k,l}]_{1\leq k,l\leq N}$ is given by
\begin{equation}
\left\{\begin{array}{l}
 \ds d^j_{k,l}=\sum_{n\in\mathbb{Z}}\tilde g_j[l-k+nN]   \\
 \ds r^j_{k,l}=\sum_{n\in\mathbb{Z}} g_j[k-l+nN]
\end{array}\right.\mbox{, for }j=0\ldots J \text{.}
\end{equation}

We assume that the considered analysis filters have unit norms; i.e., $\sum_n (d^j_{l,n})^2=1$, for $l=1\ldots N$, $j=0\ldots J$. By convention, we also assume that, for $j=1\dots J$, each $\D_j$ implements a highpass channel; i.e., $\forall l$, $\sum_n d^j_{l,n}=0$. The complementary lowpass channel is then implemented by $\D_0$ with $\forall l$, $\sum_n d^0_{l,n}=2^{J/2}$.

Hence, denoting by $\w_j=\D_j\y=[w^j_l]_{1\leq l\leq N}$ (resp. $\oomega_j=\D_j\x=[\omega^j_l]_{1\leq l\leq N}$) each vector of noisy (resp. noise-free) transform coefficients, we have
\begin{equation}
\label{eq:highpass}
\forall l=1\ldots N,
\left\{
\begin{array}{ccl}
\E{w^j_l}{}
&=
&\omega^j_l\mbox{, for }j=1\dots J\text{,}\\
\E{w^0_l}{}
&=
&\omega^0_l+2^{J/2}K \text{.}
\end{array}
\right.
\end{equation}

While the noisy highpass coefficients are unbiased estimates of their noise-free counterparts, the lowpass coefficients exhibit a constant bias $(2^{J/2}K)$ that must be removed.

\item The circulant matrix $\tD=[\tD\transp_0\:\:\tD\transp_1\ldots\tD\transp_J]\transp$ implements a linear estimation of the variance $\ttw_j=\tD_j\y=[\tw^j_l]_{1\leq l\leq N}$ of each transform coefficient $w^j_l$. The actual variance is given by
\begin{eqnarray}
\label{eq:coefVar1}
\var{w^j_l}=\sum^N_{n=1}(d^j_{l,n})^2\var{y_n}\stackrel{\eqref{eq:momentsNC2}}{=}4\sum^N_{n=1}(d^j_{l,n})^2\left(x_n+\frac{K}{2}\right)
\text{.}
\end{eqnarray}

Since $\E{y_n}=x_n+K$, the natural choice $\tD_j=[(d^j_{l,n})^2]_{1\leq l,n \leq N}$ achieves
\begin{eqnarray}
\label{eq:coefVar3}
\var{w^j_l}=4\left(\E{\tw^j_l}-\frac{K}{2}\right) \text{.}
\end{eqnarray}
\item The vector function $\ttheta:\mathbb{R}^L\times\mathbb{R}^L\rightarrow\mathbb{R}^L$, where $L=(J+1)N$, can generally be arbitrary, from a simple pointwise thresholding rule to more sophisticated multivariate processing. In this section, we will however consider only subband-adaptive \emph{pointwise} processing; i.e.,
\begin{equation}
\label{eq:SAPF}
\ttheta(\w,\ttw)=[\theta^j_l(w^j_l,\tw^j_l)]_{0\leq j \leq J,1\leq l \leq N} \text{.}
\end{equation}
\end{itemize}

We further assume that the transform-domain pointwise processing of~\eqref{eq:SAPF} is (at least) continuously differentiable, with piecewise-differentiable partial derivatives. Introducing the notation
\begin{multline*}
\qquad \partial_{1}\ttheta(\w,\ttw) = \left[\frac{\partial \theta_l(\w,\ttw)}{\partial w_l}\right]_{1\leq l \leq L},
\partial_{2}\ttheta(\w,\ttw) = \left[\frac{\partial \theta_l(\w,\ttw)}{\partial \tw_l}\right]_{1\leq l \leq L}, \\
\!\!\!\!\partial^2_{11}\ttheta(\w,\ttw) = \left[\frac{\partial^2 \theta_l(\w,\ttw)}{\partial w_l^2}\right]_{1\leq l \leq L}
\!\!\!\!,\partial^2_{22}\ttheta(\w,\ttw) = \left[\frac{\partial^2 \theta_l(\w,\ttw)}{\partial \tw_l^2}\right]_{1\leq l \leq L}
\!\!\!\!,\partial^2_{12}\ttheta(\w,\ttw) = \left[\frac{\partial^2\theta_l(\w,\ttw) }{\partial w_l\partial \tw_l}\right]_{1\leq l \leq L}
\end{multline*}
and denoting by ``$\odot$'' the Hadamard (element-wise) matrix product, we have the following.

\begin{corollary}
\label{co:TDcure}
For pointwise processing of the form given by~\eqref{eq:SAPF} and satisfying the requirements of Lemma~\ref{lem:CURE}, the risk estimate of~\eqref{eq:CURE} takes the following form:
\begin{eqnarray}
\label{eq:TDpointwiseCURE}
\emph{CURE}
&=
&\frac{1}{N}\left(\Vert \f(\y)-(\y-K\cdot\1)\Vert^2-\4\transp\left(\y-\frac{K}{2}\cdot\1\right)\right)\nonumber\\
&
&+\frac{8}{N}\left(\y-\frac{K}{2}\cdot\1\right)\transp\left((\R\odot\D\transp)\partial_1\ttheta(\w,\ttw)+(\R\odot\tD\transp)\partial_2\ttheta(\w,\ttw)\right)\nonumber\\
&
&-\frac{8}{N}\y\transp\left((\R\odot\D\transp\odot\D\transp)\partial^2_{11}\ttheta(\w,\ttw)+(\R\odot\tD\transp\odot\tD\transp)\partial^2_{22}\ttheta(\w,\ttw)\right)\nonumber\\
&
&+\frac{16}{N}\y\transp(\R\odot\D\transp\odot\tD\transp)\partial^2_{12}\ttheta(\w,\ttw) \text{.}
\end{eqnarray}
\end{corollary}
The proof of this result is straightforwardly obtained by developing the term $(\y-K/2~\cdot~\1)\transp\partial\f(\y)-\y\transp\partial^2\f(\y)$ from~\eqref{eq:CURE} for $\ttheta(\w,\ttw)$ as defined in~\eqref{eq:SAPF}. A similar result for transform-domain denoising of mixed Poisson-Gaussian data is proved in~\cite{Luisier2011a}.

For the remainder of this section, we drop the subband superscript $j$ and the in-band location index $l$. We thus denote by $w$, $\tw$, and $\omega$ any of the $w^j_l$, $\tw^j_l$, and $\omega^j_l$, for $j=1\ldots J$.

\subsection{Choice of thresholding rule}
\label{ssec:UWTthresh}
We need to specify a particular shrinkage or thresholding rule to estimate each unknown highpass coefficient $\omega$ from its noisy counterpart $w$. In the minimum-mean-squared error sense, the optimal pointwise shrinkage factor is given by
\begin{equation}
\label{eq:shrinkageMMSE}
\alpha^*=\arg\min_{\alpha}\E{(\alpha w-\omega)^2}\stackrel{\eqref{eq:highpass},\eqref{eq:coefVar3}}{=}1-\frac{4\left(\E{\tw}-\frac{K}{2}\right)}{\E{w^2}}
\text{.}
\end{equation}

There are various possible implementations of the above formula to yield an effective shrinkage function. For the case of $K=2$ degrees of freedom, Nowak proposed in~\cite{Nowak1999} the function
\begin{eqnarray}
\label{eq:shrinkageNowak}
\theta(w,\tw)=\max\Big(1-\lambda\frac{4\max(\tw-1,1)}{w^2},0\Big)w \text{,}
\end{eqnarray}
where the particular choice $\lambda=3$ was motivated by a Gaussian prior on the noisy coefficients $w$. Our experiments have indicated that replacing $\max(\tw-1,1)$ by $\tw$ gives slightly better MSE performance, and so, following the recent idea of \emph{linear expansion of thresholds} (LET)~\cite{Blu2007}, we propose the following shrinkage rule for arbitrary degrees of freedom:
\begin{eqnarray}
\label{eq:shrinkageLET}
\theta(w,\tw;\a)=\sum^I_{i=1}a_i\underbrace{\max\Big(1-\lambda_i\frac{4\tw}{w^2},0\Big)w}_{\theta_i(w,\tw)}, \quad \text{with $I<<N$}
\text{,}
\end{eqnarray}
which can be seen as an optimized generalization of~\eqref{eq:shrinkageNowak}.  To satisfy the requirements of Corollary~\ref{co:TDcure}, we implement a continuously differentiable approximation of the $\max(\cdot)$ function.

\begin{figure}[h!]
\begin{center}
\includegraphics[width=.5\columnwidth]{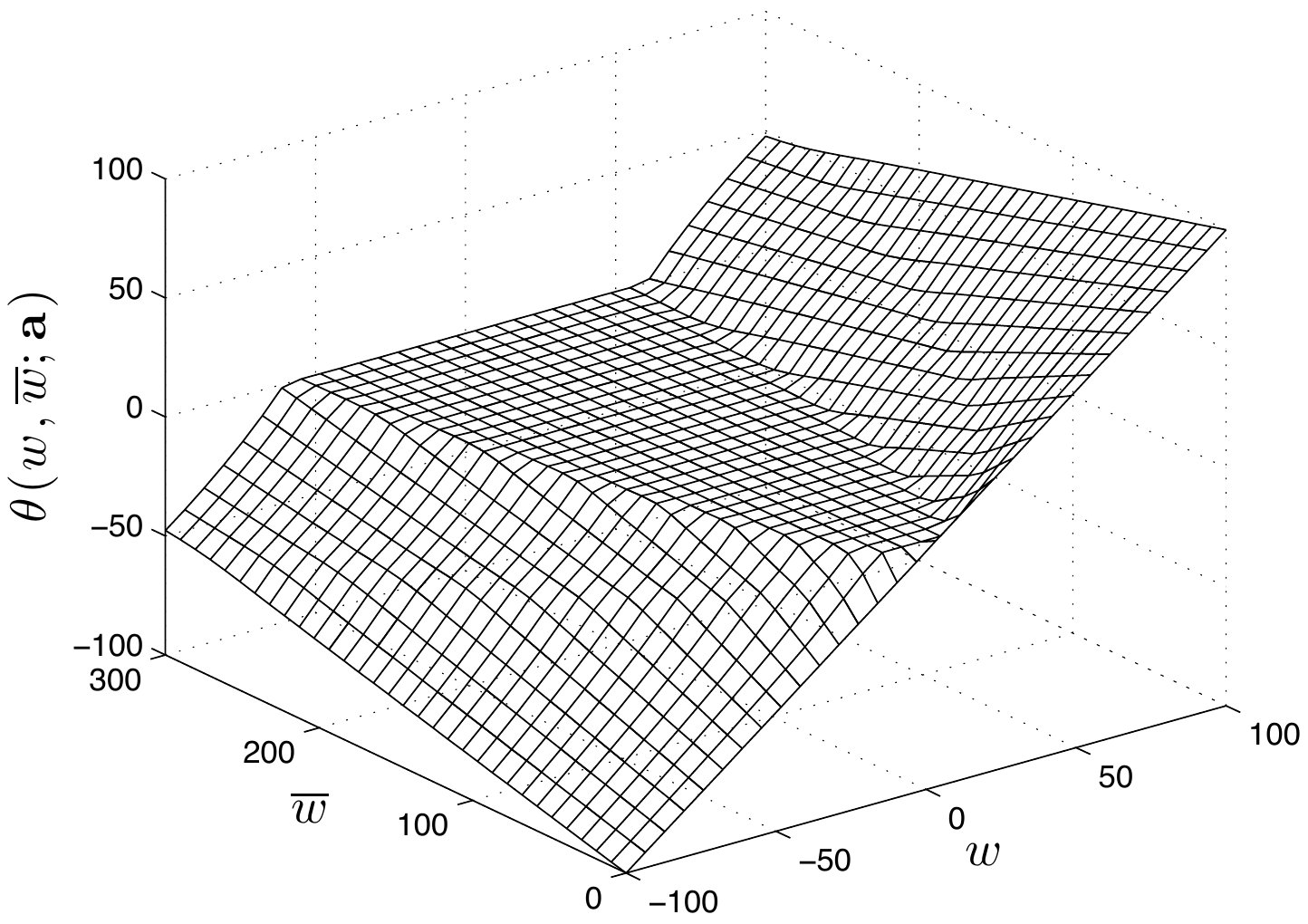}\vspace{-1em}
\caption{Possible realization of the proposed thresholding rule of~\eqref{eq:shrinkageLET} ($I=2$, $\lambda_1=3$, $\lambda_2=9$, $\a=[0.75\,0.25]\transp$).}
\label{fig:shrinkage}
\vspace{-2em}
\end{center}
\end{figure}

Empirically we have observed $I=2$ terms per subband to be the best choice in~\eqref{eq:shrinkageLET}.  The vector $\a\in\mathbb{R}^I$ of subband-adaptive parameters can be optimized in closed form via least squares, while $\lambda_1$ and $\lambda_2$ can be optimized by minimizing the risk estimate of~\eqref{eq:TDpointwiseCURE} directly. However, fixing $\lambda_1=3$ and $\lambda_2=9$ was observed to work well in all of our experiments, and leads to a much faster implementation.  (We observed values close to $(3,9)$ to yield equivalent results $\pm 0.2\dB$.)  A potential realization of the proposed LET is displayed in Fig.~\ref{fig:shrinkage}.

\subsection{Implementation}
\label{ssec:UWTimp}
The overall subband-adaptive transform-domain estimator is thus, for $\mathcal{I} = \{0\ldots J\} \times \{1,2\}$,
\begin{equation*}
\f(\y)= \sum_{\mathbf{i}\in\mathcal{I}}a_\mathbf{i}\f_\mathbf{i}(\y) = a^0_1\underbrace{(\R_0\w_0-K)}_{\parbox{5em}{\centering\scriptsize Lowpass \\*[-0.5em]bias removed}} +
\sum^J_{j=1}\sum^2_{i=1}a^j_{i}\R_j\ttheta^j_{i}(\w_j,\ttw_j)\text{,}
\end{equation*}
with the CURE-optimized parameters $\a=[a_\mathbf{i}]_{\mathbf{i}\in\mathcal{I}}$ the solution to $\M\a=\c$, where
\begin{equation}
\label{eq:LEToptim}
\left\{
\begin{array}{ccl}
\c
&=
&\ds\left[\left(\y-K\cdot\1\right)\transp\f_\mathbf{i}(\y)-4\left(\left(\y-\frac{K}{2}\cdot\1\right)\transp\partial\f_\mathbf{i}(\y)-\y\transp\partial^2\f_\mathbf{i}(\y)\right)\right]_{\mathbf{i}\in\mathcal{I}} \text{,}\\
\M
&=
&\left[\f_\mathbf{i}(\y)\transp\f_\mathbf{j}(\y)\right]_{\mathbf{i},\mathbf{j}\in\mathcal{I}} \text{.}
\end{array}
\right.
\end{equation}

\section{CURE-Optimized Denoising via Unnormalized Haar Wavelet Transform}
\label{sec:CUREhaar}
In the previous section, we have considered the general case of an undecimated filterbank transform and derived the corresponding \emph{image-domain} MSE estimate.  Owing to the intractability of the noncentral chi-square distribution after an arbitrary (even orthogonal) transformation, an explicit \emph{transform-domain} risk estimate is generally unobtainable. Remarkably, in the particular case of the \emph{unnormalized} Haar wavelet transform, the derivation of such an explicit subband-dependent MSE estimate is possible. Its construction is presented in this section.

\subsection{Unnormalized Haar wavelet-domain CURE}
\label{ssec:CUREhaar}
\begin{figure}
\begin{center}
\setlength{\unitlength}{0.7mm}
\thinlines
\begin{picture}(185,45)(5,-5)
\multiput(2,40)(4,0){46}{\line(1,0){2}}
\multiput(2,-7)(4,0){46}{\line(1,0){2}}
\multiput(2,-7)(0,4){12}{\line(0,1){2}}
\multiput(185,-7)(0,4){12}{\line(0,1){2}}

\put(5,25){\filtreA{1-z^{-1}}{2}{\,\w^j}}
\put(5,30){\line(0,-1){25}}
\put(5,0){\filtreA{1+z^{-1}}{2}{\,\s^j}}
\put(50,5){\vector(1,0){10}}
\put(50,30){\vector(1,0){10}}
\put(0,18.5){\vector(1,0){5}}
\put(0,19.5){\makebox(0,0)[r]{$\s^{j-1}$}}

\put(62,7){\line(0,1){11}}
\put(62,18){\line(1,0){30}}
\put(92,18){\vector(0,1){7}}
\put(67,30){\vector(1,0){10}}
\put(66,5){\vector(1,0){8}}
\put(77,25){\framebox(31,10){${\bm\theta}^j(\w^j,\s^j)$}}

\put(74,3){\makebox{\parbox{7em}{\footnotesize\centering Same scheme\\[-0.5em]applied recursively}}}
\multiput(74,-3)(0,4){4}{\line(0,1){2}}
\multiput(112,-3)(0,4){4}{\line(0,1){2}}
\multiput(74,-3)(4,0){10}{\line(1,0){2}}
\multiput(74,11)(4,0){10}{\line(1,0){2}}

\put(108,30){\line(1,0){12}}
\put(127,28.3){\makebox[0em][r]{$\hat{\bm\omega}^{j}$}}

\put(112,5){\line(1,0){8}}
\put(127,3.3){\makebox[0em][r]{$\hat{\bm\varsigma}^{j}$}}

\put(127,25){\filtreS{\mbox{\large$\frac{1-z}{2}$}}{2}{}}
\put(182,30){\line(0,-1){25}}
\put(127,0){\filtreS{\mbox{\large$\frac{1+z}{2}$}}{2}{}}
\put(127,5){\vector(1,0){10}}
\put(127,30){\vector(1,0){10}}
\put(182,17.5){\makebox(0,0)[c]{$\oplus$}}
\put(182,17.5){\vector(1,0){8}}
\put(191,19){\makebox(0,0)[l]{$\hat{\bm\varsigma}^{j-1}$}}
\end{picture}
\caption{Signal-dependent noise reduction in the unnormalized Haar discrete wavelet transform.}
\label{fig:unHaar}
\vspace{-2em}
\end{center}
\end{figure}
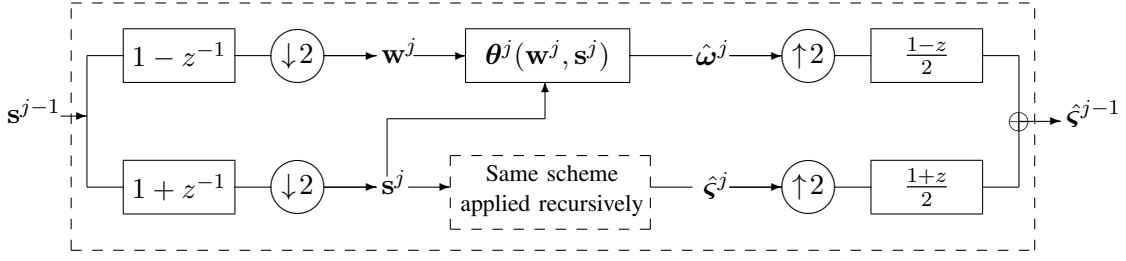

The 1D unnormalized Haar discrete wavelet transform consists of a critically-sampled two-channel filterbank (see Fig.~\ref{fig:unHaar}). On the analysis side, the lowpass (resp. highpass) channel is implemented by the unnormalized Haar scaling (resp. wavelet) filter whose $z$-transform is $\tilde{H}(z^{-1})=1+z^{-1}$ (resp. $\tilde{G}(z^{-1})=1-z^{-1}$). To achieve a perfect reconstruction, the synthesis side is implemented by the lowpass and highpass filters $H(z)=(1+z)/2$ and $G(z)=(1-z)/2$.
At a given scale $j$, the unnormalized Haar scaling and wavelet coefficients of the observed data $\y=\s^0$ are given by
\begin{equation}
\label{eq:unHaarY}
s^j_n = s^{j-1}_{2n}+s^{j-1}_{2n-1}, \quad
w^j_n = s^{j-1}_{2n}-s^{j-1}_{2n-1} \text{.}
\end{equation}

Similarly, the unnormalized Haar scaling and wavelet coefficients of the noncentrality parameter vector of interest $\x=\bm\varsigma^0$ are given by
\begin{equation}
\label{eq:unHaarX}
\varsigma^j_n = \varsigma^{j-1}_{2n}+\varsigma^{j-1}_{2n-1}, \quad
\omega^j_n = \varsigma^{j-1}_{2n}-\varsigma^{j-1}_{2n-1} \text{.}
\end{equation}

Since the sum of independent noncentral chi-square random variables is a noncentral chi-square random variable whose noncentrality parameter and number of degrees of freedom are the summed noncentrality parameters and number of degrees of freedom~\cite{Patnaik1949}, the empirical scaling coefficients follow a noncentral chi-square distribution; i.e.,
$\s^j\sim \chi^2_{K_j}(\vvarsigma^j)\mbox{, where }K_j=2^j K$.  Moreover, since the squared lowpass filter coefficients are the lowpass filter coefficients themselves, the scaling coefficients can be used as estimates of the variance of the (same-scale) wavelet coefficients. In the notation of Section~\ref{ssec:CUREuwt}, this means that $\ttw^j=\s^j$ for $j=1\ldots J$.

Denoting by $N_j$ the number of samples at a given scale $j$ and assuming a subband-adaptive processing $\ttheta^j:\mathbb{R}^{N^j}\times\mathbb{R}^{N^j}\rightarrow\mathbb{R}^{N^j}$, the MSE in each highpass subband $j$ is given by $\MSE_j=\frac{1}{N_j}\Vert\ttheta^j(\w^j,\s^j)-\oomega^j\Vert^2$, and we have the following theorem.

\begin{theorem}
\label{th:CUREj}
Let $\ttheta(\w,\s) = \ttheta^j(\w^j, \s^j)$ be an estimator of the unnormalized Haar wavelet coefficients $\oomega=\oomega^j$ of $\x$ at scale $j$, satisfying the conditions of Lemma~\ref{lem:CURE}. Then the random variable
\begin{multline}
\label{eq:CUREj}
\!\!\!\!\!\!\!\!\!\!\!\!\emph{CURE}_j
=\frac{1}{N_j}\left(\|\ttheta(\w,\s)\!-\!\w\|^2\!-\!\4\transp\left(\s\!-\!\frac{K_{j}}{2}\cdot\1\right)\right)+\frac{8}{N_j}\left(\left(\s\!-\!\frac{K_{j}}{2}\cdot\1\right)\transp\partial_{1}\ttheta(\w,\s)+\w\transp\partial_{2}\ttheta(\w,\s)\right)\\
\!-\!\frac{8}{N_j}\left(\w\transp\left(\partial^2_{11}\ttheta(\w,\s)+\partial^2_{22}\ttheta(\w,\s)\right)+2\s\transp\partial^2_{12}\ttheta(\w,\s)\right)
\end{multline}
is an unbiased estimate of the risk for subband $j$; i.e., $\ds\E{\emph{CURE}_j}= \E{\emph{MSE}_j}$.
\end{theorem}
\begin{IEEEproof}
We consider the case $j = 1$, so that we may use $K=K_j/2$, $\y = \s^{j-1}$, and $\x=\bm\varsigma^{j-1}$ to ease notation.  We first develop the squared error between $\oomega$ and its estimate $\ttheta(\w,\s)$:
\begin{equation}
\label{equ:0}
\E{\|\ttheta(\w,\s)-\oomega\|^2}= \E{\|\ttheta(\w,\s)\|^2}-2\underbrace{\E{\oomega\transp\ttheta(\w,\s)}}_{\textrm{(I)}}+\underbrace{\|\oomega\|^2}_{\textrm{(II)}}
\text{.}
\end{equation}

We can then evaluate the two expressions (I, II) that involve the unknown $\oomega$.
\begin{description}
\item (I)\quad Computation of $\ds\E{\oomega\transp\ttheta(\w,\s)}=\sum^{N_j}_{n=1}\E{\omega_n\theta_n(\w,\s)}$: We can successively write
\begin{eqnarray}
\label{eq:res1}
\E{\omega_n \theta_n(\w,\s)}
&\hspace{-1em}\stackrel{\eqref{eq:unHaarX}}{=}
&\hspace{-1em}\E{x_{2n}\theta_n(\w,\s)}{}-\E{x_{2n-1}\theta_n(\w,\s)}{}\nonumber\\
&\hspace{-1em}\stackrel{\eqref{eq:CURElem},\eqref{eq:unHaarY}}{=}	
&\hspace{-1em}\E{w_n\left(\theta_n(\w,\s)+4\left(\partial^2_{11}\theta_n(\w,\s)+\partial^2_{22}\theta_n(\w,\s)-\partial_{2}\theta_n(\w,\s)\right)\right)}\nonumber\\
&&-4\E{(s_n-K)\partial_{1}\theta_n(\w,\s)-2s_n\partial^2_{12}\theta_n(\w,\s)}
\text{.}
\end{eqnarray}
\item (II)\quad Computation of $\ds\|\oomega\|^2=\sum^{N_j}_{n=1}\omega^2_n$: We can successively write
\begin{equation}
\label{eq:res2}
\omega^2_n=\E{\omega_n w_n}\stackrel{\eqref{eq:res1}}{=}\E{w^2_n-4(s_n-K)}
\text{.}
\end{equation}
\end{description}

Inserting~\eqref{eq:res1},~\eqref{eq:res2} into~\eqref{equ:0} yields the desired equality; for $j>1$, the proof is similar.
\end{IEEEproof}

Subband superscript $j$ will be omitted below, as we consider any of the $J$ wavelet subbands.
\subsection{CUREshrink}
\label{ssec:CUREshrink}
A natural choice of subband-adaptive estimator in orthogonal wavelet representations is \emph{soft thresholding}, introduced by Weaver \etal\cite{Weaver1991} and theoretically justified by Donoho~\cite{Donoho1995b}. In contrast to the AWGN scenario, a signal-dependent threshold is required here. As in the case of Poisson noise removal~\cite{Hirakawa2009a,Luisier2010a}, we wish to adapt the original ``uniform'' soft thresholding as
\begin{equation}
\label{eq:CUREshrink}
\theta_n(\w,\s;a)=\sign(w_n)\max(|w_n|-a\sqrt{s_n},0) \text{.}
\end{equation}

\begin{figure}[h!]
\begin{center}
\begin{tabular}{cc}
\includegraphics[width=.4\columnwidth]{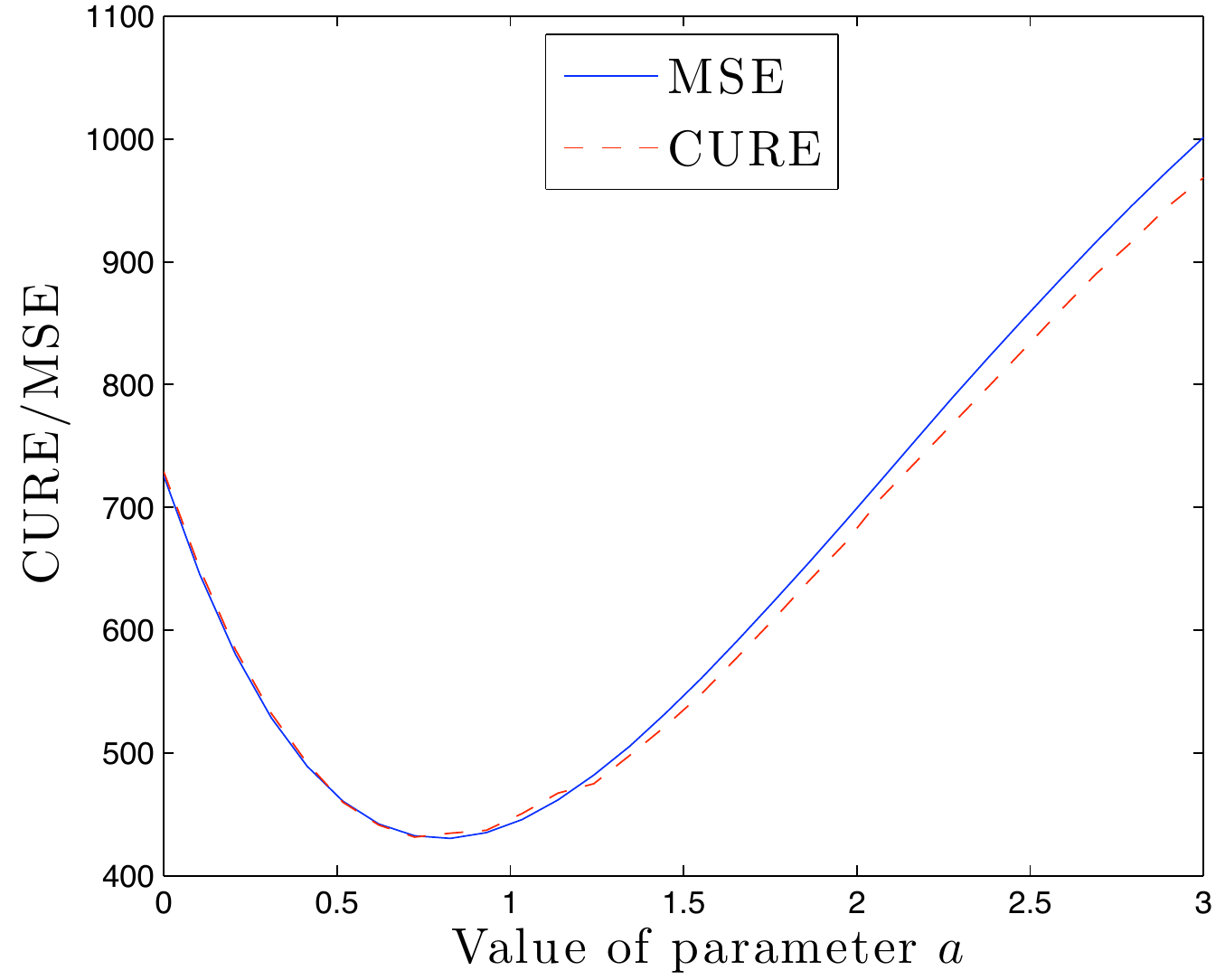}
&\includegraphics[width=.4\columnwidth]{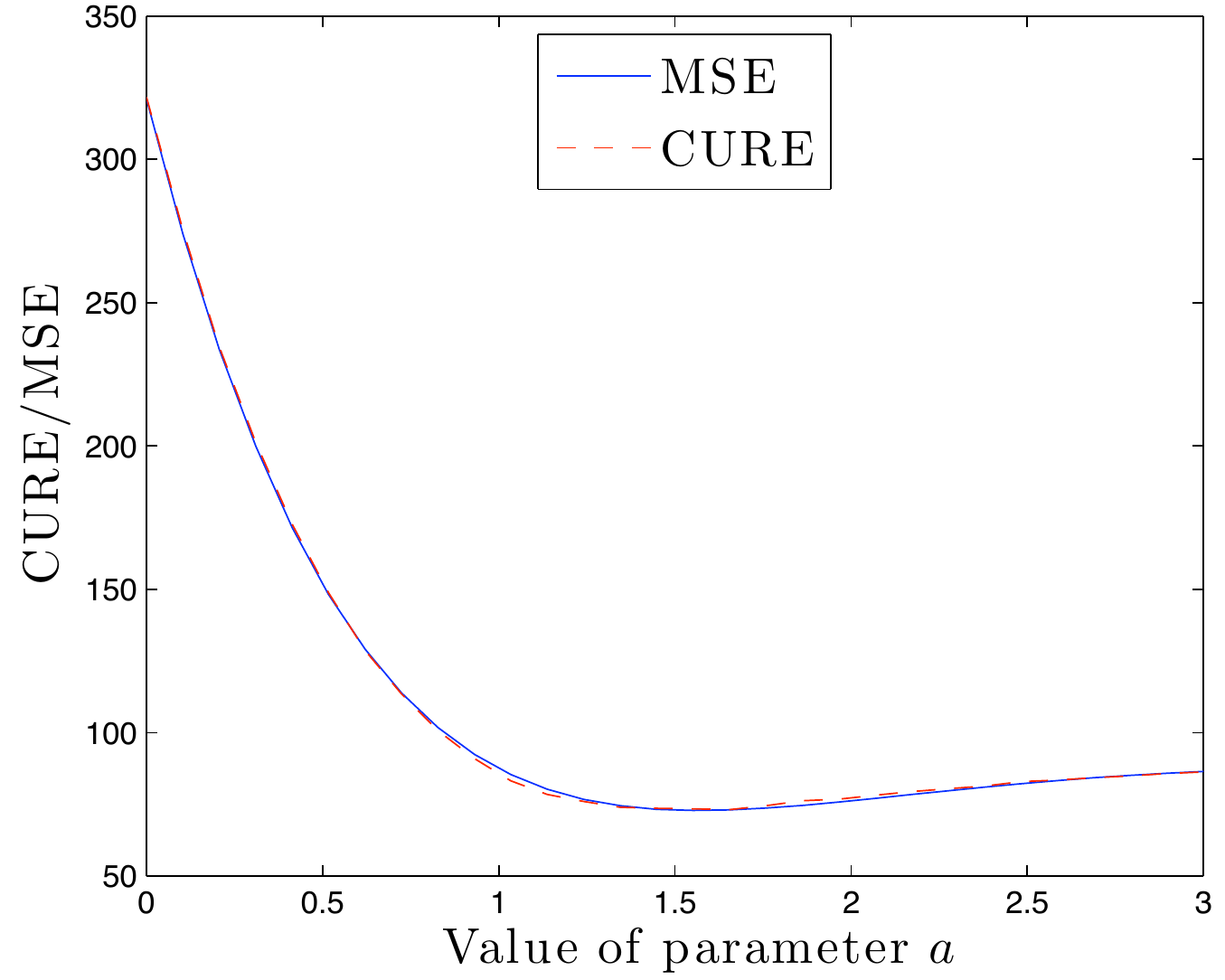}
\end{tabular}
\caption{Minimum CURE vs.~MMSE threshold selection for the CUREshrink adapted soft thresholding of~\eqref{eq:CUREshrink}. Left: $N=128\times128$ samples, $\s\sim \chi^2_{8}(\vvarsigma)$, input $\SNR=15\dB$. Right: $N=256\times256$, $\s\sim \chi^2_{16}(\vvarsigma)$, input $\SNR=10\dB$.}
\label{fig:CUREvsMSE}
\vspace{-2em}
\end{center}
\end{figure}

In \emph{SUREshrink} and \emph{PUREshrink}~\cite{Donoho1995,Hirakawa2009a,Luisier2010a} for Gaussian (resp. Poisson) noise reduction, $a$ is set to the value that minimizes the corresponding unbiased risk estimate. Similarly, we may select $a$ to yield the minimum CURE value according to~\eqref{eq:CUREj} on the basis of observed data $\y$, resulting in a \emph{CUREshrink} denoising procedure. To comply with the requirements of Theorem~\ref{th:CUREj}, we use a continuously differentiable approximation to soft thresholding. Figure~\ref{fig:CUREvsMSE} shows the empirical accuracy of CURE as a practical criterion for choosing the best value of $a$; we have also observed a pointwise LET approach, as in~\eqref{eq:shrinkageLET}, to yield comparable denoising results.

\subsection{Joint inter-/intra-scale CURE-LET}
\label{ssec:CURE-LET}
To decrease the usual ringing artifacts inherent to orthogonal transform-domain thresholding, more sophisticated denoising functions must be considered. In particular, the integration of inter-scale dependencies between wavelet coefficients (the so-called ``parent-child'' relationship) has already been shown to significantly increase the denoising quality in both AWGN reduction~\cite{Vetterli2000b,Sendur2002a,Portilla2003,Luisier2007a} and Poisson intensity estimation~\cite{Luisier2010a}.

\begin{figure}[h!]
\begin{center}
\begin{tabular}{cc}
\includegraphics[width=.4\columnwidth]{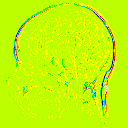}
&\includegraphics[width=.4\columnwidth]{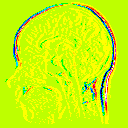}
\end{tabular}
\caption{A child (left) and its group-delay compensated parent (right) in a particular subband at the first scale of a 2D unnormalized Haar wavelet transform. In this 16-color map, the background yellowish value is zero, with the most significant coefficients appearing either in blue (negative) or in red (positive).}
\label{fig:parent}
\vspace{-2em}
\end{center}
\end{figure}

To this end, Fig.~\ref{fig:parent} shows an example of a group-delay compensated parent $\p$ and its child $\w$ for a particular subband at the first scale of a 2D unnormalized Haar wavelet transform.
For the unnormalized Haar wavelet transform, the group-delay compensated parent $\p=[p_n]_{1\leq n\leq N}$ is simply given by $p_n=s_{n+1}-s_{n-1}
$~\cite{Luisier2010a}.  From Fig.~\ref{fig:parent}, we observe that both the signs and the locations of the significant (i.e., large-magnitude) coefficients persist across scale.  To take advantage of this persistence, we propose the following LET approach, inspired by~\eqref{eq:shrinkageLET}:
\begin{multline}
\label{eq:multiLET}
\theta_n(\w,\s;\a) =
\sum^2_{k=1}a_k\max\Big(1-\lambda_k\frac{4 \gamma_n(\s)}{\gamma^2_n(\w)},0\Big)w_n+
\sum^2_{k=1}a_{k+2}\max\Big(1-\lambda_k\frac{4 \gamma_n(\s)}{\gamma^2_n(\p)},0\Big)w_n\\
+\sum^2_{k=1}a_{k+4}\max\Big(1-\lambda_k\frac{4 \gamma_n(\s)}{\gamma^2_n(\w)},0\Big)p_n+
\sum^2_{k=1}a_{k+6}\max\Big(1-\lambda_k\frac{4 \gamma_n(\s)}{\gamma^2_n(\p)},0\Big)p_n
\text{.}
\end{multline}
Here the function $\gamma_n(\u)=1/\sqrt{2\pi}\sum_k|u_k|\mathrm{e}^{-(n-k)^2/2}$ implements a normalized Gaussian smoothing of the magnitude of its argument; this local filtering accounts for similarities between neighboring wavelet, scaling, and parent coefficients.

The proposed denoising function of~\eqref{eq:multiLET} thus integrates both the inter- and intra-scale dependencies that naturally arise in the Haar wavelet transform. It involves a set $\a$ of eight parameters that can be optimized via least squares, as well as parameters $\lambda_1$ and $\lambda_2$ that can be fixed in advance without noticeable loss in denoising quality; we use $\lambda_1=1$ and $\lambda_2=9$ in all experiments below.   Considering the processing of all wavelet coefficients in a given subband $j$,~\eqref{eq:multiLET} reads as $\ttheta(\w,\s;\a)=\sum^8_{k=1}a_k\ttheta_k(\w,\s)$. The optimal (in the minimum CURE sense) set of linear parameters is then the solution of the linear system of equation $\M\a=\c$, where
\begin{equation*}
\left\{
\begin{array}{ccl}
\c
&=
&\ds\Big[\w\transp\ttheta_k(\w,\s)-4\left(\s-\frac{K_j}{2}\cdot\1\right)\transp\partial_1\ttheta_k(\w,\s)+8\s\transp\partial^2_{12}\ttheta_k(\w,\s)+\\
&&4\w\transp\left(\partial^2_{11}\ttheta_k(\w,\s)+\partial^2_{22}\ttheta_k(\w,\s)-\partial_2\ttheta_k(\w,\s)\right)\Big)\Big]_{1\leq k\leq 8}\text{,}\\
\M
&=
&\left[\ttheta_k(\w,\s)\transp\ttheta_l(\w,\s)\right]_{1\leq k,l\leq 8} \text{.}
\end{array}
\right.
\end{equation*}
Finally, the bias is removed from the lowpass residual subband at scale $J$ as $\hat{{\bs\varsigma}}^J = \s^J-2^JK\cdot\1$.

\section{Application to Magnitude MR Image Denoising}
\label{sec:Experiments}
In magnitude magnetic resonance imaging, the observed image consists of the magnitudes $|m_n|$ of $N$ independent complex measurements $m_n$, where
\begin{equation}
\left\{
\begin{array}{ccc}
\re{m_n}
&\sim
&\mathcal{N}(\re{\mu_n},\sigma^2) \text{,}\\
\im{m_n}
&\sim
&\mathcal{N}(\im{\mu_n},\sigma^2) \text{.}
\end{array}
\right.
\end{equation}

Our objective is to estimate the original (unknown) magnitudes $|\mu_n|=\sqrt{\re{\mu_n}^2+\im{\mu_n}^2}$ from their noisy observations $|m_n|$. If we define two $N$-dimensional vectors
\begin{equation}
\label{eq:rescale}
\left\{
\begin{array}{ccc}
\x
&=
&[|\mu_n|^2/\sigma^2]_{1\leq n\leq N}\in\mathbb{R}^N_{+} \text{,}\\
\y
&=
&[|m_n|^2/\sigma^2]_{1\leq n\leq N}\in\mathbb{R}^N_{+} \text{,}
\end{array}
\right.
\end{equation}
then the data likelihood for $\y$ is the product of $N$ independent noncentral $\chi^2$ distributions with $K=2$ degrees of freedom and noncentrality parameter $x_n$; i.e.,~\eqref{eq:NC2} with $K=2$.

We then denoise the magnitude MR image $\m$ according to the following steps:
\begin{enumerate}
\item If necessary, estimate the noise variance $\sigma^2$ using known techniques;
\item Rescale the squared-magnitude MR image $\y$ according to~\eqref{eq:rescale};
\item Apply a CURE-optimized denoising algorithm to obtain an estimate $\hx=\f(\y)$ of $\x$;
\item Fix some $\lambda\in[0,1]$ and produce the final estimate $\hat{\mmu}$ of the unknown magnitude MR image $\mmu$, by applying the following nonlinear rescaling function:
\begin{equation}
\label{eq:finalestimate}
\hat{\mmu}=\sigma\left[\lambda\sqrt{|f_n(\y)|}+(1-\lambda)\sqrt{\max(f_n(\y),0)}\right]_{1\leq n\leq N}
\text{.}
\end{equation}
\end{enumerate}
We evaluate denoising performance objectively using three full-reference image quality metrics:
\begin{enumerate}
\item The standard peak signal-to-noise ratio (PSNR), defined as $ \PSNR=10\log_{10}\frac{N\Vert\mmu\Vert^2_\infty}{\Vert\hat{\mmu}-\mmu\Vert^2}$;
\item A contrast-invariant PSNR, defined as $\CIPSNR=10\log_{10}\frac{N\Vert\mmu\Vert^2_\infty}{\Vert (a^{*}\hat{\mmu}+b^{*})-\mmu\Vert^2}$, where the affine parameters $a^{*},b^{*}$ are given by
\begin{equation*}
(a^{*},b^{*})=\arg\min_{a,b}\Vert (a\hat{\mmu}+b)-\mmu\Vert^2\Longleftrightarrow
\left\{
\begin{array}{ccl}
a^{*}
&=
&\ds\frac{N\mmu\transp\hat{\mmu}-\1\transp\mmu\:\1\transp\hat{\mmu}}{N\hat{\mmu}\transp\hat{\mmu}-(\1\transp\hat{\mmu})^2}\text{,}\smallskip\\
b^{*}
&=
&\ds\frac{\1\transp\mmu\:\hat{\mmu}\transp\hat{\mmu}-\mmu\transp\hat{\mmu}\:\1\transp\hat{\mmu}}{N\hat{\mmu}\transp\hat{\mmu}-(\1\transp\hat{\mmu})^2}
\text{.}
\end{array}
\right.
\end{equation*}
\item The mean of the structural similarity index map (SSIM), a popular visual quality metric introduced in~\cite{Wang2004} (see \url{https://ece.uwaterloo.ca/~z70wang/research/ssim/}).
\end{enumerate}

In all simulated experiments, we have assumed that the variance $\sigma^2$ of the complex Gaussian noise is known. In practice, a reliable estimate can be obtained in signal-free or constant regions of the image by moment matching~\cite{Nowak1999,Fernandez2008b} or maximum likelihood techniques~\cite{Sijbers1998b,He2009,Sijbers2007}. When no background is available, more sophisticated approaches can be considered~\cite{Rajan2010}.

\begin{figure}[h!]
\begin{center}
\begin{tabular}{cccc}
\hspace{-0.75em}\textbf{\emph{Image 1}} $\bs{(256\times256)}$
&\hspace{-0.75em}\textbf{\emph{Image 2}} $\bs{(512\times512)}$
&\hspace{-0.75em}\textbf{\emph{Image 3}} $\bs{(256\times256)}$
&\hspace{-0.75em}\textbf{\emph{Image 4}} $\bs{(256\times256)}$\\
\hspace{-0.75em}\includegraphics[scale=0.4]{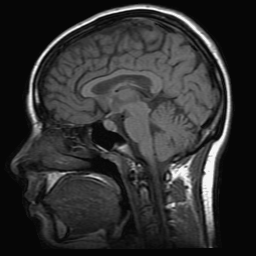}
&\hspace{-0.75em}\includegraphics[scale=0.2]{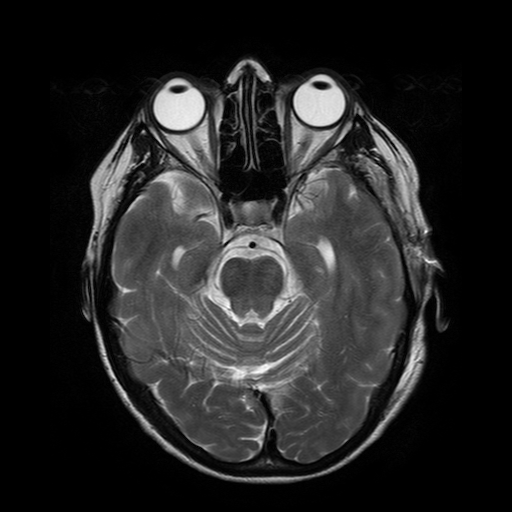}
&\hspace{-0.75em}\includegraphics[scale=0.4]{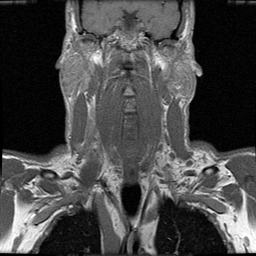}
&\hspace{-0.75em}\includegraphics[scale=0.4]{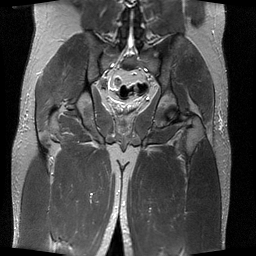}
\end{tabular}
\caption{Test set of high-quality magnitude MR images used in the experiments of Section~\ref{sec:Experiments}.}
\label{fig:TestImages}
\vspace{-3em}
\end{center}
\end{figure}

To simulate various input noise levels, several values for $\sigma$ have been selected in the range $\sigma\in[5,100]$. The set of high-quality magnitude MR test images used is shown in Fig.~\ref{fig:TestImages}, and may be obtained from \url{http://bigwww.epfl.ch/luisier/MRIdenoising/TestImages.zip}.

\begin{figure}[h!]
\begin{center}
\begin{tabular}{cc}
\textbf{\emph{Image 1}}
&\textbf{\emph{Image 2}}\\
\includegraphics[scale=0.4]{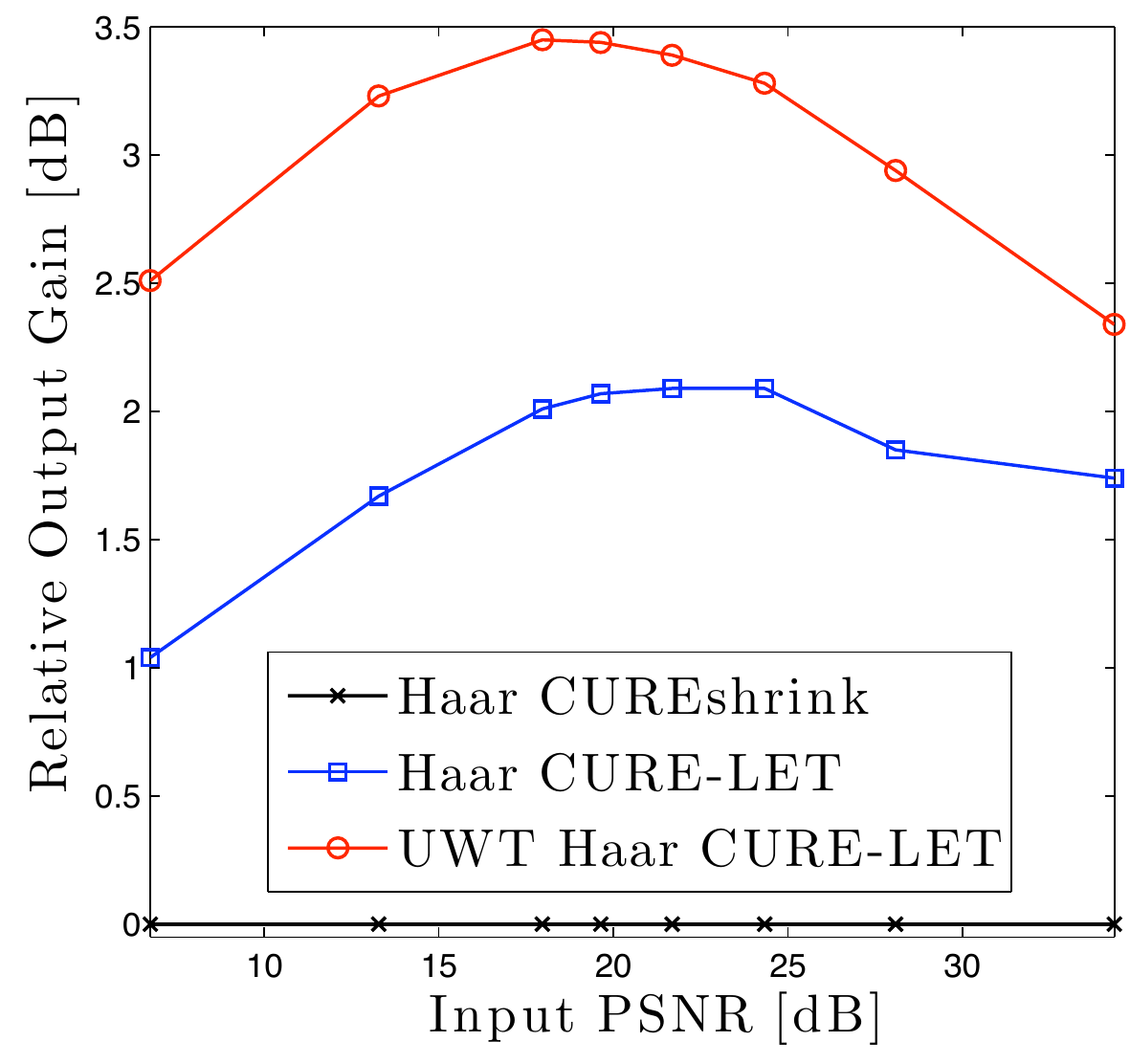}
&\includegraphics[scale=0.4]{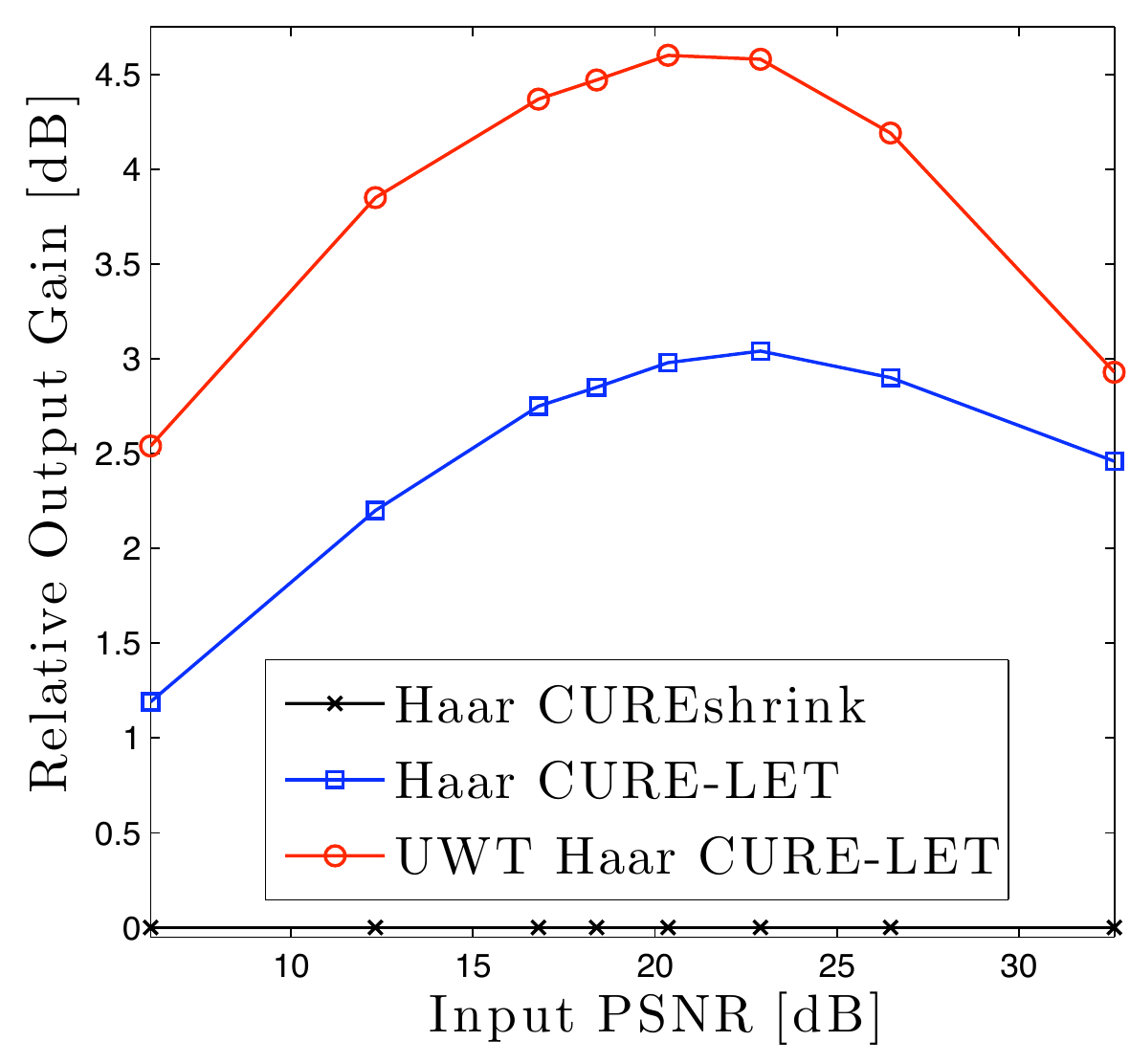}\vspace{-1em}
\end{tabular}
\caption{PSNR comparisons among pointwise undecimated Haar CURE-LET (Section~\ref{ssec:UWTthresh}) and joint intra-/inter-scale unnormalized Haar CURE-LET (Section~\ref{ssec:CURE-LET}), shown relative to unnormalized Haar CUREshrink (Section~\ref{ssec:CUREshrink}).}
\label{fig:shrinkVSlet}
\vspace{-2em}
\end{center}
\end{figure}
\begin{figure}[h!]
\begin{center}
\begin{tabular}{cc}
\textbf{\emph{Image 1}}
&\textbf{\emph{Image 3}}\\
\includegraphics[scale=0.4]{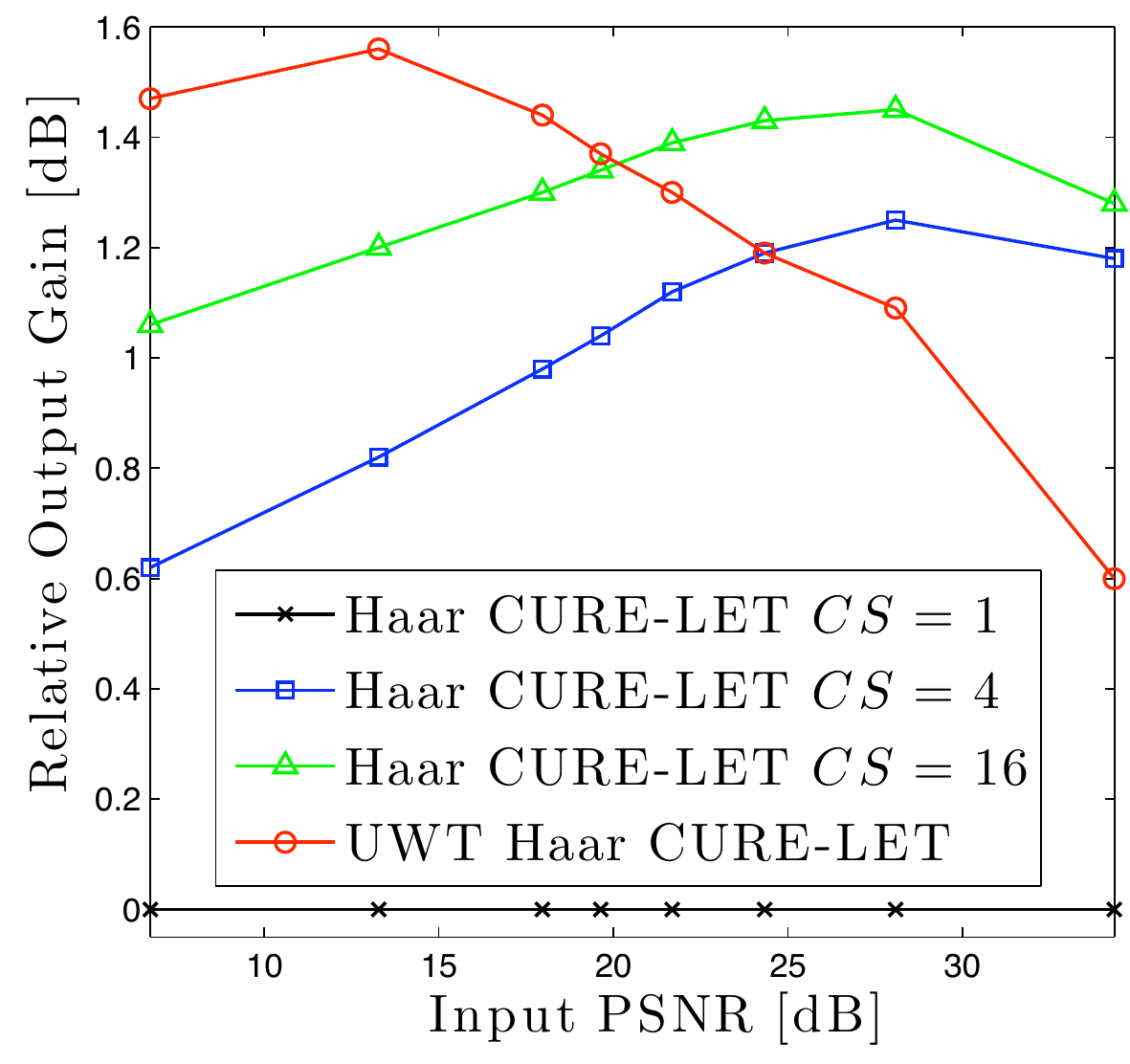}
&\includegraphics[scale=0.4]{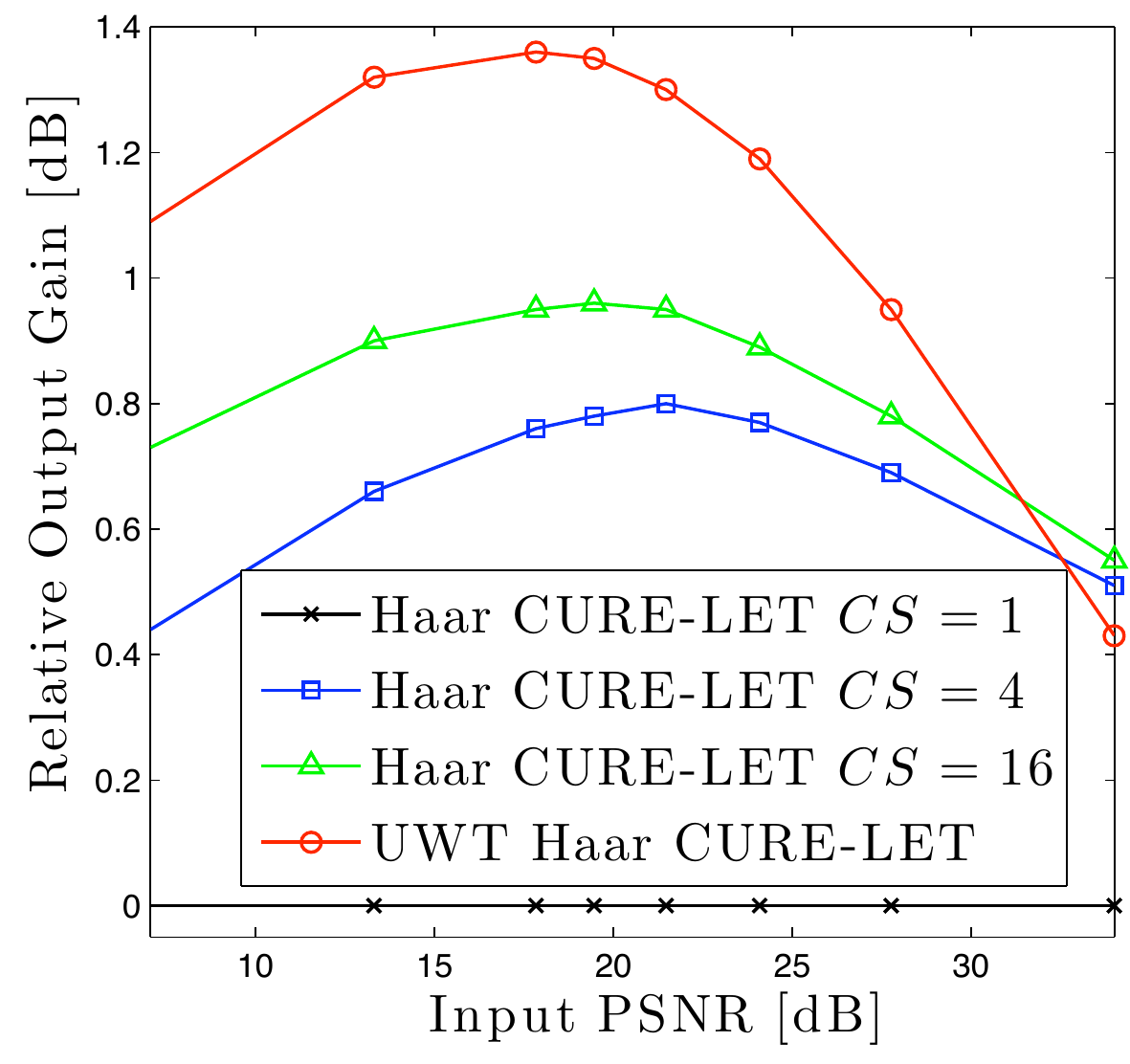}\vspace{-1em}
\end{tabular}
\caption{PSNR improvements brought by cycle-spinning the unnormalized Haar wavelet transform.}
\label{fig:cs}
\vspace{-2em}
\end{center}
\end{figure}
\begin{figure}[h!]
\begin{center}
\begin{tabular}{cc}
\textbf{\emph{Image 1}}
&\textbf{\emph{Image 2}}\\
\includegraphics[scale=0.4]{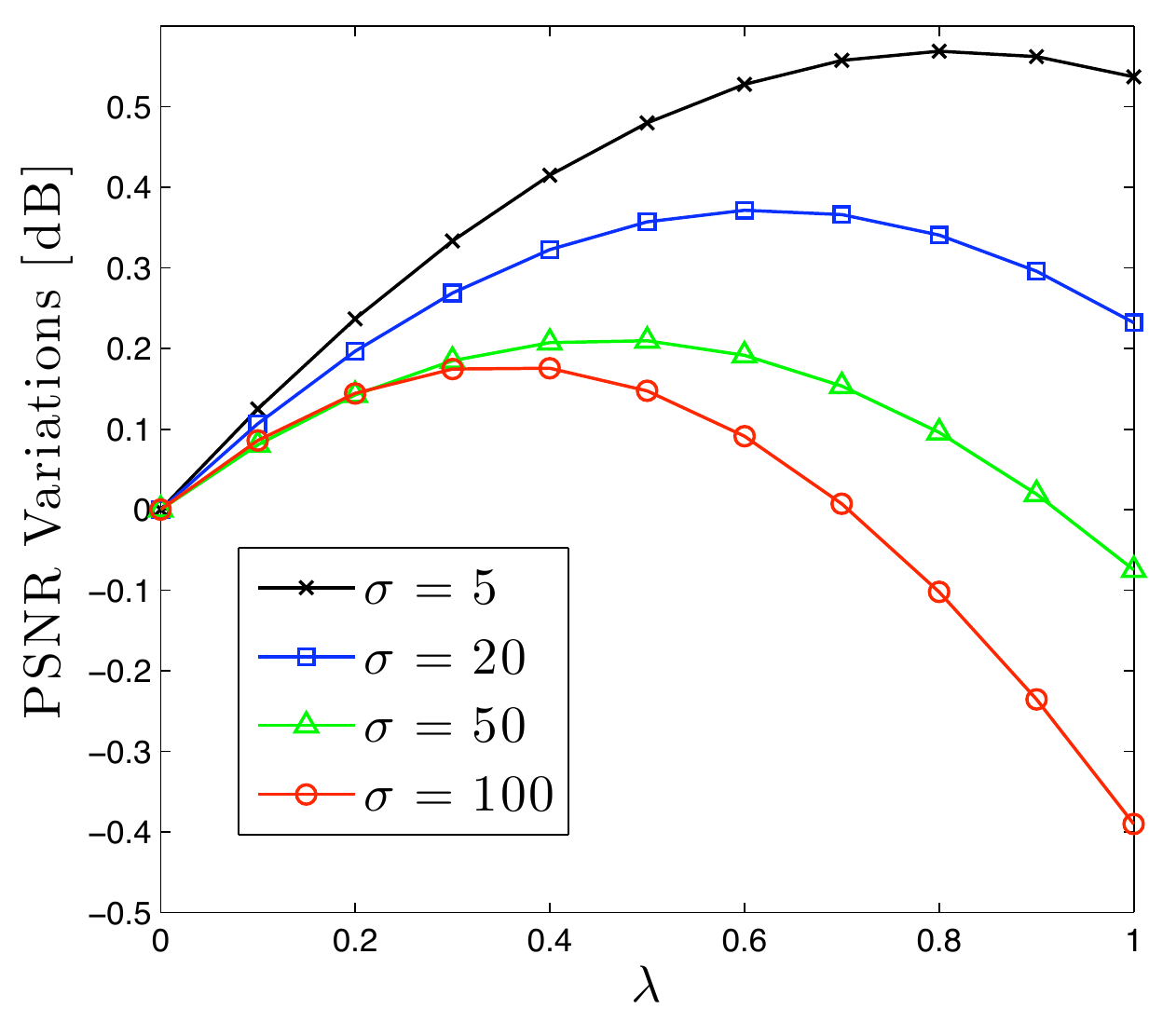}
&\includegraphics[scale=0.4]{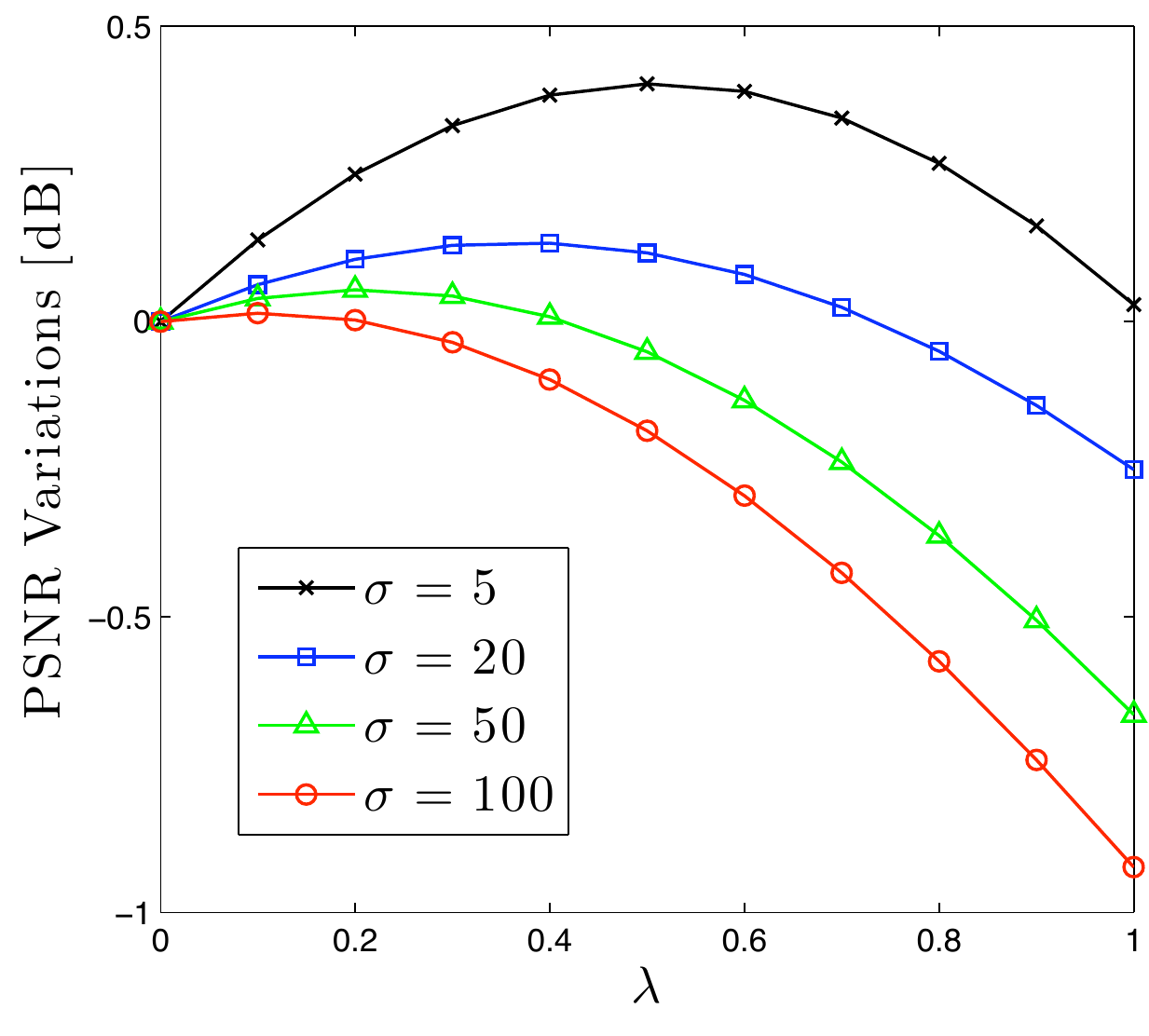}\vspace{-1em}
\end{tabular}
\caption{Sensitivity to the choice of $\lambda$ in~\eqref{eq:finalestimate}; ordinate shows PSNR variations relative to the reference case $\lambda=0$.}
\label{fig:lambda}
\vspace{-3em}
\end{center}
\end{figure}

\subsection{Comparisons between variants of the proposed approach}
Before comparing our approach with four state-of-the-art MR image denoising methods, we first evaluate the performance of the various variants of our approach. In Fig.~\ref{fig:shrinkVSlet}, we compare the results obtained using the pointwise CURE-LET thresholding of~\eqref{eq:shrinkageLET}, applied in the undecimated Haar wavelet transform domain, and the joint intra-/inter-scale LET denoising function of~\eqref{eq:multiLET}, applied to the unnormalized Haar wavelet transform coefficients.  These are shown relative to a baseline provided by the unnormalized Haar  CUREshrink approach of~\eqref{eq:CUREshrink}.  As expected, the joint intra-/inter-scale denoising function of~\eqref{eq:multiLET} outperforms the simple soft-thresholding of~\eqref{eq:CUREshrink} by $1-3\dB$.  Moreover, pointwise thresholding applied in a shift-invariant setting outperforms (by $0.5-1.5\dB$) a more sophisticated thresholding in a shift-variant one.

Note that the shift-invariance of the unnormalized Haar wavelet transform can be increased by applying the so-called ``cycle-spinning'' technique~\cite{Coifman1995}. In Fig.~\ref{fig:cs}, we show the PSNR improvements brought by averaging the results of several cycle-spins (CS). As observed, 16 cycle-spins allow a near match in performance to the shift-invariant transform. The cycle-spinning technique has the further advantage of being easily implementable in parallel.

In Fig.~\ref{fig:lambda}, we evaluate the sensitivity of the CURE-optimized algorithms with respect to the value of parameter $\lambda$ appearing in the final nonlinear reconstruction function of~\eqref{eq:finalestimate}. Note that this parameter selection is usually not addressed in the literature; the most common choices are either $\lambda=0$ as in~\cite{Manjon2008} or $\lambda=1$ as in~\cite{Pizurica2003}. Yet, Fig.~\ref{fig:lambda} indicates that the MMSE choice usually lies in between these two extremal values. In all our experiments, we have thus used $\lambda=0.5$.

\subsection{Comparisons to state-of-the-art MR image denoising methods}
As benchmarks for evaluating our CURE-LET approach, we have retained four state-of-the-art MR image denoising techniques: two wavelet-based algorithms~\cite{Nowak1999,Pizurica2003} (code at \url{http://telin.ugent.be/~sanja/}), a spatially adaptive linear MMSE filter~\cite{Fernandez2008b}, and an unbiased nonlocal means filter specifically designed for MR data~\cite{Manjon2008} (code at \url{http://personales.upv.es/jmanjon/denoising/nlm2d.htm}). For each of these methods, we have used the tuning parameters suggested in their respective publications and software, except for the linear MMSE filter, where we have hand-optimized (in the MMSE sense) the size of the filter support.

We have considered three CURE-LET variants: 16 cycle-spins of the joint intra-/inter-scale thresholding of~\eqref{eq:multiLET} applied in the unnormalized Haar wavelet transform; the pointwise thresholding of~\eqref{eq:shrinkageLET} applied in the undecimated Haar wavelet transform; and the same pointwise thresholding applied in a mixed-basis overcomplete transform (an undecimated Haar wavelet transform plus an $8\times 8$ overlapping block discrete cosine transform, BDCT). Note that a LET spanning several overcomplete bases was previously considered in~\cite{Luisier2011a}.

In Table~\ref{tab:PSNRcomp}, the PSNR results of the various methods are displayed, while the contrast-invariant CIPSNR results are reported in Table~\ref{tab:CIPSNRcomp}. As observed, the proposed CURE-optimized pointwise LET applied in a mixed-basis overcomplete transform consistently achieves the best results. The average denoising gains are about $+2~\dB$ relative to the method of~\cite{Nowak1999}, $+3~\dB$ compared to~\cite{Pizurica2003}, $+3~\dB$ compared to~\cite{Fernandez2008b}, and $+0.4~\dB$ compared to~\cite{Manjon2008}.  Overall, these CURE-LET methods compare very favorably to state-of-the-art approaches, especially under very noisy conditions, in which case the signal-dependent nature of the noise is more pronounced. Note that further denoising gains are likely to be obtained by considering more sophisticated thresholding rules and image-adaptive overcomplete dictionaries (e.g.,~\cite{Aharon2006}).
\begin{table}[h!]
\centering
\caption{PSNR Comparisons}\vspace{-1.5em}
\scriptsize
\begin{tabular}{||c||c|c|c|c|c|c||c|c|c|c|c|c||}
\hline
$\sigma$
&5
&10
&20
&30
&50
&100
&5
&10
&20
&30
&50
&100
\\
\hline
\hline
\textbf{Image}
&\multicolumn{6}{c||}{\textbf{\textit{Image 1}} $\bs{256\times256}$}
&\multicolumn{6}{c||}{\textbf{\textit{Image 2}} $\bs{512\times512}$}\\
\hline
\textbf{Input PSNR}
&34.35
&28.09
&21.69
&17.97
&13.28
&6.74
&32.62
&26.48
&20.36
&16.80
&12.32
&6.14
\\
\hline
\hline
\cite{Nowak1999}
&37.95
&33.46
&29.09
&26.51
&23.20
&18.35
&38.72
&34.13
&29.41
&26.55
&22.85
&17.83
\\
\hline
\cite{Pizurica2003}
&33.13
&30.47
&28.56
&25.13
&18.56
&10.48
&33.79
&32.17
&26.64
&22.00
&16.55
&9.47
\\
\hline
\cite{Fernandez2008b}
&37.29
&32.86
&28.10
&25.22
&21.69
&17.33
&37.28
&32.19
&27.31
&24.58
&21.37
&17.64
\\
\hline
\cite{Manjon2008}
&39.15
&34.97
&30.65
&28.15
&24.79
&19.04
&39.85
&35.98
&31.72
&28.89
&25.08
&19.61
\\
\hline
\begin{minipage}{1.7cm}
\centering
CURE-LET\\*[-0.8em]
(Haar, CS=16)
\end{minipage}
&39.00
&34.84
&30.71
&28.15
&24.87
&20.32
&40.02
&35.88
&31.60
&28.95
&25.50
&20.86
\\
\hline
\begin{minipage}{1.7cm}
\centering
UWT Haar\\*[-0.8em]
CURE-LET
\end{minipage}
&38.32
&34.48
&30.62
&28.29
&25.23
&20.73
&39.42
&35.95
&32.07
&29.53
&26.27
&21.56
\\
\hline
\begin{minipage}{1.7cm}
\centering
UWT/BDCT\\*[-0.8em]
CURE-LET
\end{minipage}
&\textbf{39.19}
&\textbf{35.00}
&\textbf{30.89}
&\textbf{28.50}
&\textbf{25.38}
&\textbf{21.02}
&\textbf{40.06}
&\textbf{36.14}
&\textbf{32.24}
&\textbf{29.75}
&\textbf{26.37}
&\textbf{21.60}
\\
\hline
\hline
\textbf{Image}
&\multicolumn{6}{c||}{\textbf{\textit{Image 3}} $\bs{256\times256}$}
&\multicolumn{6}{c||}{\textbf{\textit{Image 4}} $\bs{256\times256}$}\\
\hline
\textbf{Input PSNR}
&34.00
&27.77
&21.48
&17.84
&13.32
&7.07
&33.84
&27.73
&21.64
&18.13
&13.71
&7.36
\\
\hline
\hline
\cite{Nowak1999}
&35.88
&31.72
&27.64
&25.21
&22.10
&17.84
&35.75
&31.50
&27.36
&24.99
&22.07
&18.08
\\
\hline
\cite{Pizurica2003}
&29.35
&27.62
&25.89
&24.73
&19.45
&11.65
&28.20
&26.32
&24.63
&24.15
&20.98
&12.31
\\
\hline
\cite{Fernandez2008b}
&35.84
&31.24
&26.58
&24.07
&21.00
&17.47
&35.86
&31.19
&26.70
&24.14
&21.06
&17.40
\\
\hline
\cite{Manjon2008}
&36.25
&32.51
&28.92
&26.67
&23.53
&19.17
&36.16
&32.24
&28.55
&26.43
&23.43
&18.68
\\
\hline
\begin{minipage}{1.7cm}
\centering
CURE-LET\\*[-0.8em]
(Haar, CS=16)
\end{minipage}
&36.55
&32.54
&28.72
&26.47
&23.57
&19.67
&36.36
&32.17
&28.23
&25.98
&23.19
&19.41
\\
\hline
\begin{minipage}{1.7cm}
\centering
UWT Haar\\*[-0.8em]
CURE-LET
\end{minipage}
&36.43
&32.71
&29.07
&26.88
&23.99
&20.03
&36.20
&32.24
&28.58
&26.49
&23.77
&20.00
\\
\hline
\begin{minipage}{1.7cm}
\centering
UWT/BDCT\\*[-0.8em]
CURE-LET
\end{minipage}
&\textbf{36.88}
&\textbf{33.02}
&\textbf{29.24}
&\textbf{27.01}
&\textbf{24.10}
&\textbf{20.09}
&\textbf{36.56}
&\textbf{32.51}
&\textbf{28.72}
&\textbf{26.59}
&\textbf{23.89}
&\textbf{20.12}
\\
\hline
\end{tabular}\\*[0.5em]
\begin{minipage}[h]{1\columnwidth}
\centering
\footnotesize
Output PSNRs have been averaged over 10 noise realizations.
\end{minipage}
\vspace{-3em}
\label{tab:PSNRcomp}
\end{table}

\begin{table}[h!]
\centering
\caption{CIPSNR Comparisons}\vspace{-1.5em}
\scriptsize
\begin{tabular}{||c||c|c|c|c|c|c||c|c|c|c|c|c||}
\hline
$\sigma$
&5
&10
&20
&30
&50
&100
&5
&10
&20
&30
&50
&100
\\
\hline
\hline
\textbf{Image}
&\multicolumn{6}{c||}{\textbf{\textit{Image 1}} $\bs{256\times256}$}
&\multicolumn{6}{c||}{\textbf{\textit{Image 2}} $\bs{512\times512}$}\\
\hline
\textbf{Input CIPSNR}
&34.52
&28.58
&22.72
&19.69
&16.85
&15.16
&34.71
&28.66
&22.59
&19.29
&16.08
&14.13
\\
\hline
\hline
\cite{Nowak1999}
&38.00
&33.57
&29.41
&27.05
&24.02
&19.46
&39.35
&35.15
&31.02
&28.50
&25.13
&20.27
\\
\hline
\cite{Pizurica2003}
&33.55
&30.89
&28.64
&26.29
&22.89
&18.72
&36.41
&33.36
&28.63
&25.88
&22.81
&18.94
\\
\hline
\cite{Fernandez2008b}
&37.30
&32.91
&28.33
&25.57
&22.17
&17.91
&37.81
&33.18
&28.57
&26.06
&23.01
&19.29
\\
\hline
\cite{Manjon2008}
&39.15
&34.99
&30.66
&28.16
&24.80
&19.21
&39.93
&36.11
&31.94
&29.17
&25.42
&19.92
\\
\hline
\begin{minipage}{1.7cm}
\centering
CURE-LET\\*[-0.8em]
(Haar, CS=16)
\end{minipage}
&39.00
&34.85
&30.82
&28.40
&25.30
&20.75
&40.29
&36.37
&32.50
&30.13
&26.98
&22.46
\\
\hline
\begin{minipage}{1.7cm}
\centering
UWT Haar\\*[-0.8em]
CURE-LET
\end{minipage}
&38.32
&34.48
&30.64
&28.36
&25.41
&21.03
&39.56
&36.17
&32.45
&30.07
&27.08
&22.71
\\
\hline
\begin{minipage}{1.7cm}
\centering
UWT/BDCT\\*[-0.8em]
CURE-LET
\end{minipage}
&\textbf{39.20}
&\textbf{35.01}
&\textbf{30.91}
&\textbf{28.58}
&\textbf{25.60}
&\textbf{21.36}
&\textbf{40.30}
&\textbf{36.45}
&\textbf{32.68}
&\textbf{30.34}
&\textbf{27.22}
&\textbf{22.79}
\\
\hline
\hline
\textbf{Image}
&\multicolumn{6}{c||}{\textbf{\textit{Image 3}} $\bs{256\times256}$}
&\multicolumn{6}{c||}{\textbf{\textit{Image 4}} $\bs{256\times256}$}\\
\hline
\textbf{Input CIPSNR}
&34.33
&28.31
&22.24
&18.88
&15.44
&13.06
&34.07
&28.01
&22.13
&19.06
&16.07
&14.07
\\
\hline
\hline
\cite{Nowak1999}
&36.01
&31.88
&28.00
&25.76
&22.90
&18.83
&35.94
&31.77
&27.79
&25.51
&22.64
&18.53
\\
\hline
\cite{Pizurica2003}
&30.25
&28.49
&26.35
&24.98
&22.02
&18.13
&28.96
&26.93
&25.07
&24.46
&22.34
&18.25
\\
\hline
\cite{Fernandez2008b}
&35.86
&31.38
&26.88
&24.58
&21.65
&18.17
&35.93
&31.35
&26.93
&24.41
&21.34
&17.68
\\
\hline
\cite{Manjon2008}
&36.27
&32.51
&28.93
&26.67
&23.56
&19.21
&36.17
&32.26
&28.57
&26.46
&23.48
&18.72
\\
\hline
\begin{minipage}{1.7cm}
\centering
CURE-LET\\*[-0.8em]
(Haar, CS=16)
\end{minipage}
&36.58
&32.59
&28.86
&26.73
&24.03
&20.29
&36.42
&32.27
&28.44
&26.27
&23.55
&19.67
\\
\hline
\begin{minipage}{1.7cm}
\centering
UWT Haar\\*[-0.8em]
CURE-LET
\end{minipage}
&36.48
&32.74
&29.12
&26.97
&24.19
&20.44
&36.25
&32.28
&28.66
&26.61
&23.96
&20.24
\\
\hline
\begin{minipage}{1.7cm}
\centering
UWT/BDCT\\*[-0.8em]
CURE-LET
\end{minipage}
&\textbf{36.93}
&\textbf{33.07}
&\textbf{29.29}
&\textbf{27.11}
&\textbf{24.31}
&\textbf{20.55}
&\textbf{36.61}
&\textbf{32.56}
&\textbf{28.82}
&\textbf{26.73}
&\textbf{24.09}
&\textbf{20.42}
\\
\hline
\end{tabular}\\*[0.5em]
\begin{minipage}[h]{1\columnwidth}
\centering
\footnotesize
Output CIPSNRs have been averaged over 10 noise realizations.
\end{minipage}
\label{tab:CIPSNRcomp}
\vspace{-4.3em}
\end{table}

\begin{figure}[h!]
\begin{center}
\begin{tabular}{ccc}
\hspace{-1em}(A)
&\hspace{-0.5em}(B)
&\hspace{-0.5em}(C)\\
\hspace{-1em}\includegraphics[width=0.33\columnwidth]{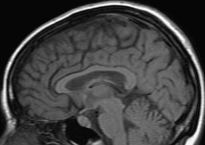}
&\hspace{-0.5em}\includegraphics[width=0.33\columnwidth]{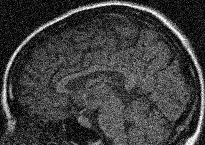}
&\hspace{-0.5em}\includegraphics[width=0.33\columnwidth]{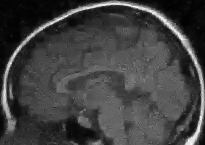}\\*[-0.75em]
\hspace{-1em}(D)
&\hspace{-0.5em}(E)
&\hspace{-0.5em}(F)\\
\hspace{-1em}\includegraphics[width=0.33\columnwidth]{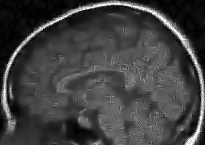}
&\hspace{-0.5em}\includegraphics[width=0.33\columnwidth]{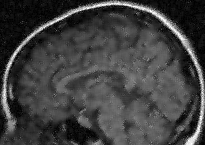}
&\hspace{-0.5em}\includegraphics[width=0.33\columnwidth]{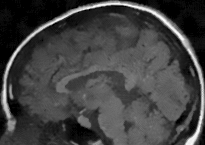}\\*[-0.75em]
\end{tabular}
\begin{tabular}{cc}
\hspace{0.25em}(G)
&\hspace{-0.5em}(H)\\
\hspace{0.25em}\includegraphics[width=0.33\columnwidth]{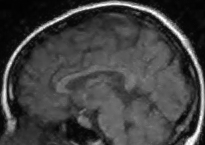}
&\hspace{-0.5em}\includegraphics[width=0.33\columnwidth]{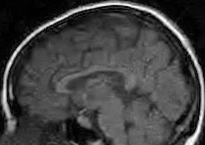}
\end{tabular}
\caption{Visual Comparisons. (A)~Zoom in image 1. (B)~Noisy version: $\SSIM~=~0.319$. (C)~Denoised by~\cite{Nowak1999}: $\SSIM~=~0.734$. (D)~Denoised by~\cite{Pizurica2003}: $\SSIM~=~0.726$. (E)~Denoised by~\cite{Fernandez2008b}: $\SSIM~=~0.653$. (F)~Denoised by~\cite{Manjon2008}: $\SSIM~=~0.719$. (G)~Denoised by Haar CURE-LET (CS=16): $\SSIM~=~0.800$. (H)~Denoised by UWT/BDCT CURE-LET: $\SSIM~=~0.796$.}
\label{fig:VisuComp}
\vspace{-3em}
\end{center}
\end{figure}

Fig.~\ref{fig:VisuComp} presents a visual comparison of the various algorithms considered. As observed, the two CURE-LET denoising results offer a good balance between noise suppression, reasonable denoising artifacts, and fine structure preservation. This subjective observation is confirmed by the higher SSIM scores obtained by the proposed denoising approach.

In Table~\ref{tab:computationTime}, we report the computation time of the various algorithms considered. All have been executed on Matlab R2010a running under Mac OS X equipped with a 2.66GHz Intel Core 2 Duo processor. As observed, the proposed CURE-LET algorithms are quite fast, taking 1--10~s to denoise a $256\times256$ image. When applied within an undecimated filterbank transform, most of the computational load is dedicated to the independent reconstructions of the processed subbands and their corresponding first and second order derivatives (see~\eqref{eq:LEToptim}).
\begin{table}[!t]
\centering
\caption{Computation times for MR image denoising techniques.}\vspace{-1.5em}
\begin{tabular}{||c||c|c||c||c|c||}
\hline
\raisebox{-1em}{\textbf{Method}}
&\multicolumn{2}{c||}{\textbf{Image size}}
&\raisebox{-1em}{\textbf{Method}}
&\multicolumn{2}{c||}{\textbf{Image size}}\\
\cline{2-3}
\cline{5-6}
&$\bs{256\times256}$
&$\bs{512\times512}$
&	
&$\bs{256\times256}$
&$\bs{512\times512}$\\
\hline
\hline
\cite{Nowak1999}
&0.2
&0.9
&Haar CURE-LET (CS=1)
&0.3
&0.8\\
\hline
\cite{Pizurica2003}
&0.6
&2.5
&Haar CURE-LET (CS=16)
&3.7
&12.1\\
\hline
\cite{Fernandez2008b}
&0.1
&0.5
&UWT CURE-LET
&1.6
&9.6\\
\hline
\cite{Manjon2008}
&55.8
&248.4
&UWT/BDCT CURE-LET
&9.7
&43.9\\
\hline						
\end{tabular}\\*[0.5em]
\begin{minipage}[h]{1\columnwidth}
\centering
\footnotesize
Computation times have been averaged over 10 runs; \cite{Fernandez2008b,Manjon2008} do not use pre-compiled MEX files.
\end{minipage}
\label{tab:computationTime}
\vspace{-3em}
\end{table}

\subsection{Denoising of a magnitude MR knee image}
We have also applied our CURE-LET denoising algorithms to an actual magnitude MR image of the knee. This $512\times512$ 16-bit raw image has been acquired on a Siemens 1.5 Tesla Magnetom Sonata MR system, following a sagittal T2-weighted protocol. The standard deviation of the complex Gaussian noise has been estimated from a signal-free region $\mathcal{S}$ of the squared data, as $\hat{\sigma}=\sqrt{\frac{1}{2}\sum_{n\in \mathcal{S}} |m_n|^2}$, and subsequently treated as known.

\begin{figure}[h!]
\begin{center}
\begin{tabular}{cc}
\textbf{(A)}
&\textbf{(B)}\\
\includegraphics[scale=0.3]{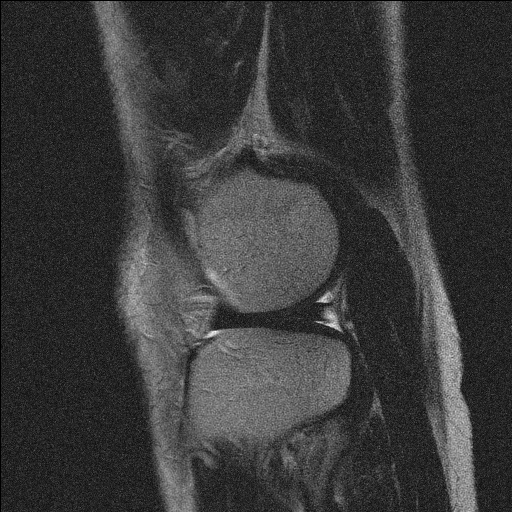}
&\includegraphics[scale=0.3]{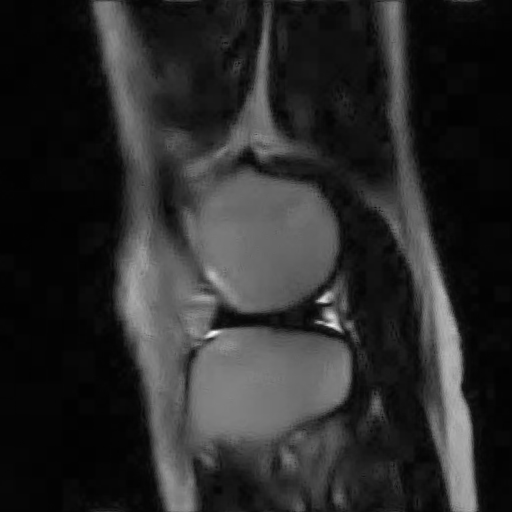}\\*[-0.75em]
\textbf{(C)}
&\textbf{(D)}\\
\includegraphics[scale=0.3]{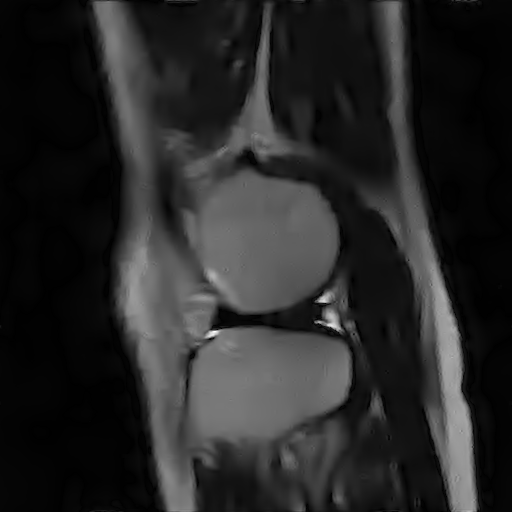}
&\includegraphics[scale=0.3]{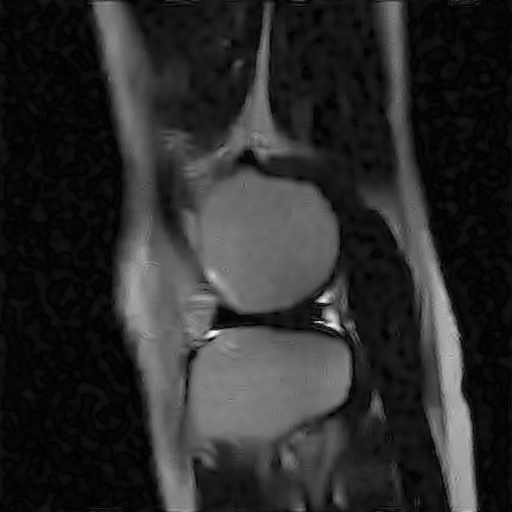}
\end{tabular}
\caption{Denoising of a magnitude MR image of the knee. (A)~Raw 16-bit data: $\hat{\sigma}=67.61$. (B)~Denoised via Haar CURE-LET (16 cycle spins). (C)~Via UWT Haar CURE-LET. (D)~Via UWT/BDCT CURE-LET.}
\label{fig:realData}
\end{center}
\end{figure}

Fig.~\ref{fig:realData} shows the denoising results of the various CURE-LET algorithms. As observed, the noise is efficiently attenuated and the contrast is significantly improved, owing to a proper reduction of the signal-dependent bias introduced by the noise.

\section{Conclusion}
\label{sec:Conclusion}

In this article we have derived a noncentral \emph{chi-square unbiased risk estimate} (CURE), and applied it to the problem of magnitude MR image denoising, where the squared value of each pixel comprises an independent noncentral chi-square variate on two degrees of freedom.  Our approach can be used to optimize the parameters of essentially any continuously differentiable estimator for this class of problems, and here we have focused our attention on transform-domain algorithms in particular.

In this vein, we first developed a  pointwise \emph{linear expansion of thresholds} (LET) estimator applied to the coefficients of an arbitrary undecimated filterbank transform.   We then considered the specific case of the unnormalized Haar wavelet transform, a multiscale orthogonal transform allowing for the derivation of \emph{subband-dependent} CURE denoising strategies.  We also introduced a subband-adaptive joint inter-/intra-scale LET that outperforms a simpler estimator similar to soft thresholding.

We then applied our proposed CURE-optimized algorithms to test images artificially degraded by noise, and observed them to compare favorably with state-of-the-art techniques, both quantitatively and qualitatively.  Finally, we showed an example of denoising results obtained on an actual magnitude MR image, in order to show the practical efficacy of our approach to MR image denoising via chi-square unbiased risk estimation.

\section{Acknowledgments}
The authors would like to thank Prof. W.-Y.~I.~Tseng for providing MRI data, and A.~Pi\v{z}urica and J.~Manj\'{o}n for making their respective software implementations available online.
\bibliographystyle{IEEEtran}
\bibliography{LuisierBluWolfe2011a}

\end{document}